\documentclass{llncs}
\pdfoutput=1
\usepackage{graphicx}
\usepackage{wrapfig}
\usepackage{amssymb}
\usepackage{amsmath}
\usepackage{multirow}
\usepackage{listings}
\usepackage{color}
\usepackage{booktabs}
\usepackage{tikz}
\usetikzlibrary{arrows,automata,shapes,shapes.multipart,decorations.markings,positioning}

\newcounter{sncolumncounter}
\newcounter{snrowcounter}

\def \nodelabel#1{%
\setcounter{snrowcounter}{1}
 \foreach \i in {#1}{%
   \draw (\sncolwidth*\value{sncolumncounter},\value{snrowcounter}) node[anchor=south]{\scriptsize\i};
   \addtocounter{snrowcounter}{1}
 }
 \addtocounter{snrowcounter}{-1}
 \addtocounter{sncolumncounter}{1}
}

\newcommand{\sncolwidth}{0.7} %

\newcommand{\addcomparator}[3][black]{%
    \draw[line width=1pt,color=#1] (\sncolwidth*\value{sncolumncounter},#2) node[circle,fill=#1,minimum size=5pt,inner sep=0pt,outer sep=0pt]{}--(\sncolwidth*\value{sncolumncounter},#3) node[circle,fill=#1,minimum size=5pt,inner sep=0pt,outer sep=0pt]{};
}

\def \addlayer{%
  \addtocounter{sncolumncounter}{1}
}

\def \nextlayer{%
  \draw [dashed] (\sncolwidth*\value{sncolumncounter}+\sncolwidth,0.6)--(\sncolwidth*\value{sncolumncounter}+\sncolwidth,\value{snrowcounter}+0.6);
  \addtocounter{sncolumncounter}{2}
}

\newenvironment{sortingnetwork}[2]
{
  \setcounter{sncolumncounter}{0}
  \setcounter{snrowcounter}{#1}
  \def \sn@fullsize{15}
  \begin{tikzpicture}[scale=#2*0.7]
}
{
  \foreach \i in {1, ..., \value{snrowcounter}}
  {
    \draw[line width=1pt] (-\sncolwidth,\i)--(\sncolwidth*\value{sncolumncounter}+\sncolwidth,\i);
  }
  \end{tikzpicture}
}
\makeatother

\begin{document}

\title{Applying Sorting Networks to Synthesize Optimized Sorting Libraries%
  \thanks{Supported by the Israel Science Foundation, grant 182/13 and
    by the Danish Council for Independent Research, Natural
    Sciences.}}

\author{Michael Codish\inst{1} \and
        Lu\'{i}s Cruz-Filipe\inst{2} \and \\
        Markus Nebel\inst{2} \and
        Peter Schneider-Kamp\inst{2}}
\authorrunning{M.~Codish, L.~Cruz-Filipe, M.~Nebel, and P.~Schneider-Kamp}
\institute{
Department of Computer Science, Ben-Gurion University of the Negev, Israel
\and
Dept. Mathematics and Computer Science, Univ. of Southern Denmark,Denmark\\
}

\maketitle
\setcounter{footnote}{0}

\begin{abstract}
  This paper presents an application of the theory of sorting networks to
  facilitate the synthesis of optimized general-purpose sorting
  libraries. Standard sorting libraries are often based on
  combinations of the classic Quicksort algorithm with insertion sort
  applied as base case for small, fixed, numbers of
  inputs. Unrolling the code for the base case by ignoring loop
  conditions eliminates branching, resulting in code
  equivalent to a sorting network. This enables %
  further program transformations based on sorting network
  optimizations, and eventually the synthesis of code from sorting
  networks.
  We show that, if considering the number of comparisons and swaps,
  the theory predicts no real advantage of this
  approach. However, significant speed-ups are obtained when taking
  advantage of instruction level parallelism and non-branching
  conditional assignment instructions, both of which are common in
  modern CPU architectures.
  We provide empirical evidence that using code synthesized from
  efficient sorting networks as the base case for Quicksort libraries
  results in significant real-world speed-ups.
\end{abstract}

\section{Introduction}

General-purpose sorting algorithms are based on comparing, and
possibly exchanging, pairs of inputs. If the order of these
comparisons is predetermined by the number of inputs to sort and does
not depend on their concrete values, then the algorithm is said to be
data-oblivious.  Such algorithms are well suited for e.g.\ parallel
sorting or secure multi-party computations.

Sorting functions in state-of-the-art programming language libraries
(such as the GNU C Library) are typically based on a variant of
Quicksort, where the base cases of the recursion apply insertion
sort:
once the subsequence to sort considered by
Quicksort falls under a certain length $M$, it is sorted
using insertion sort. The reasons for using such base cases is that,
both theoretically and empirically, insertion sort is faster than
Quicksort for sorting
small numbers of elements.
Typical values of $M$ %
are $4$ (e.g.~in the GNU C library) or $8$.

Generalizing this construction, we can take any sorting algorithm
based on the divide-and-conquer approach (e.g. Quicksort, merge sort),
and use another sorting method once the number of elements to sort in
one partition does not exceed a pre-defined limit $M$. The guiding
idea here is that, by supplying optimized code for sorting up to $M$
inputs, the overall performance of the sorting algorithm can be
improved.
One obvious way to supply optimized code for sorting up to $M$ inputs
is to provide a unique optimized implementaton of sorting $m$ elements,
for each $m \leq M$.

This approach leads directly to the following problem:
 \emph{For a given fixed number $M$, how can we obtain an
    efficient way to sort $M$ elements on a modern CPU?}
Similar questions have been asked since the 1950s, though obviously
with a different notion of what constitutes a modern CPU.

Sorting networks are a classical model of comparison-based sorting
that provides a framework for addressing such questions. In a sorting network,
$n$ inputs are fed into $n$ channels,
connected pairwise by comparators.
Each comparator compares the two
inputs from its two channels,
and outputs them sorted
back to the same two channels.
Consecutive comparators can be viewed
as a ``parallel layer'' if no two touch the same channel.
Sorting networks are data-oblivious algorithms, as the sequence of
comparisons performed is independent of the actual input.
For this
reason, they are typically viewed as hardware-oriented algorithms,
where data-obliviousness is a requirement and a fixed number of inputs
is given.

In this paper, we examine how the theory of sorting networks can
improve the performance of general-purpose software sorting
algorithms.  We show that replacing the insertion sort base case of a
Quicksort implementation as found in standard C libraries by optimized
code synthesized from logical descriptions of sorting networks leads
to significant improvements in execution times.

The idea of using sorting networks to guide the synthesis of optimized
code for base cases of sorting algorithms may seem rather obvious,
and, indeed, has been pursued earlier.
A straightforward attempt, described in~\cite{Lopez2014},
has not resulted in significant improvements, though.
In this paper we show that this is not unexpected, providing
theoretical and empirical insight into the reasons for these rather
discouraging results. In a nutshell, we provide an average case
analysis of the complexity w.r.t.\ measures such as number of
comparisons and number of swaps. From the complexity point of view,
code synthesized from sorting networks can be expected to perform
slightly worse than unrolled insertion sort.
Fortunately, for small
numbers (asymptotic) complexity arguments are not always a good
predictor of real-world performance.

The approach taken in \cite{Furtak2007} matches the advantages
of sorting networks with the vectorization instruction sets available in some modern CPU architectures.
The authors obtain significant speedups by implementing parallel comparators as vector operations,
but they require a complex heuristic algorithm to generate sequences of bit shuffling code
that needs to be executed between comparators.
Their approach is also not fully general, as they target a particular architecture.

In this paper, we combine the best of both these attempts by providing a straightforward
implementation of sorting networks that still takes advantage of the features of modern CPU architectures, while keeping generality.
We obtain speedups comparable to \cite{Furtak2007}, but our requirements to the instruction set are satisfied by virtually all modern CPUs, including those without vector operations.
The success of our approach is based on two observations.%

\begin{itemize}
\item
Sorting networks are data-oblivious and the order of comparisons is
fully determined at compile time, i.e., they are free of any
control-flow branching. Comparators can also be implemented without
branching, and on modern CPU architectures even efficiently so.
\item 
Sorting networks are inherently parallel, i.e., comparators at the
same level can be performed in parallel. Conveniently, this maps
directly to implicit \emph{instruction level parallelism} (ILP) common in
modern CPU architectures. This feature allows parallel execution of several
instructions on a single thread of a single core, as long
as they are working on disjoint sets of registers.
\end{itemize}

Avoiding branching and exploiting ILP are tasks also performed through
program transformations by the optimization stages of modern C
compilers, e.g., by unrolling loops and reordering instructions to
minimize data-dependence between neighbouring instructions. They are
though both restricted by the data-dependencies of the algorithms
being compiled and, consequently, of only limited use for
data-dependent sorting algorithms, like insertion sort.

Throughout this paper, for empirical evaluations we run all code on
an Intel Core i7, measuring runtime in CPU cycles using the time
stamp counter register using the RDTSC instruction.
As a compiler for all benchmarks, we used LLVM 6.1.0 with clang-602.0.49 as frontend on Max OS X 10.10.2.
We also tried GCC 4.8.2 on Ubuntu with Linux kernel 3.13.0-36, yielding comparable results.

The remainder of the paper is organized as follows.
Section~\ref{sec:background}
provides background information and formal
definitions for both sorting algorithms and hardware
features.
In Section~\ref{sec:complexity}, we theoretically compare Quicksort and the best known
sorting networks w.r.t.\ numbers of comparisons and swaps.
We aggressively unroll insertion sort
until we obtain a sorting network in Section~\ref{sec:unrolling}, and
in Section~\ref{sec:implementing} we show how to implement individual comparators
efficiently.
We empirically evaluate our contribution as a base case of Quicksort in Section~\ref{sec:quicksort},
before concluding and giving an outlook on future work in Section~\ref{sec:conclusion}.

\section{Background}
\label{sec:background}

\subsection{Quicksort with Insertion Sort for Base Case}
\label{subsec:background1}

For decades, Quicksort has been used in practice, due to its efficiency
in the average case. Since its first publication by Hoare~\cite{hoare},
several modifications were suggested to improve it further.
Examples are the clever choice of the pivot, or the use of a
different sorting algorithm, e.g., insertion sort, for small
subproblem sizes.  Most such suggestions have in common that the
empirically observed efficiency can be explained on theoretical
grounds by analyzing the expected number of comparisons, swaps, and
partitioning stages (see \cite{sedgewickflajolet} for details).

\begin{figure}[t]
\centering
\resizebox{\textwidth}{!}{
\setlength{\unitlength}{0.240900pt}
\ifx\plotpoint\undefined\newsavebox{\plotpoint}\fi
\sbox{\plotpoint}{\rule[-0.200pt]{0.400pt}{0.400pt}}%
\begin{picture}(1930,850)(130,50)
\sbox{\plotpoint}{\rule[-0.200pt]{0.400pt}{0.400pt}}%
\put(210.0,82.0){\rule[-0.200pt]{4.818pt}{0.400pt}}
\put(190,82){\makebox(0,0)[r]{$0$}}
\put(2018.0,82.0){\rule[-0.200pt]{4.818pt}{0.400pt}}
\put(210.0,179.0){\rule[-0.200pt]{4.818pt}{0.400pt}}
\put(190,179){\makebox(0,0)[r]{$200$}}
\put(2018.0,179.0){\rule[-0.200pt]{4.818pt}{0.400pt}}
\put(210.0,276.0){\rule[-0.200pt]{4.818pt}{0.400pt}}
\put(190,276){\makebox(0,0)[r]{$400$}}
\put(2018.0,276.0){\rule[-0.200pt]{4.818pt}{0.400pt}}
\put(210.0,373.0){\rule[-0.200pt]{4.818pt}{0.400pt}}
\put(190,373){\makebox(0,0)[r]{$600$}}
\put(2018.0,373.0){\rule[-0.200pt]{4.818pt}{0.400pt}}
\put(210.0,471.0){\rule[-0.200pt]{4.818pt}{0.400pt}}
\put(190,471){\makebox(0,0)[r]{$800$}}
\put(2018.0,471.0){\rule[-0.200pt]{4.818pt}{0.400pt}}
\put(210.0,568.0){\rule[-0.200pt]{4.818pt}{0.400pt}}
\put(190,568){\makebox(0,0)[r]{$1000$}}
\put(2018.0,568.0){\rule[-0.200pt]{4.818pt}{0.400pt}}
\put(210.0,665.0){\rule[-0.200pt]{4.818pt}{0.400pt}}
\put(190,665){\makebox(0,0)[r]{$1200$}}
\put(2018.0,665.0){\rule[-0.200pt]{4.818pt}{0.400pt}}
\put(210.0,762.0){\rule[-0.200pt]{4.818pt}{0.400pt}}
\put(190,762){\makebox(0,0)[r]{$1400$}}
\put(2018.0,762.0){\rule[-0.200pt]{4.818pt}{0.400pt}}
\put(210.0,859.0){\rule[-0.200pt]{4.818pt}{0.400pt}}
\put(190,859){\makebox(0,0)[r]{$1600$}}
\put(2018.0,859.0){\rule[-0.200pt]{4.818pt}{0.400pt}}
\put(341.0,82.0){\rule[-0.200pt]{0.400pt}{4.818pt}}
\put(341,41){\makebox(0,0){$2$}}
\put(341.0,839.0){\rule[-0.200pt]{0.400pt}{4.818pt}}
\put(602.0,82.0){\rule[-0.200pt]{0.400pt}{4.818pt}}
\put(602,41){\makebox(0,0){$4$}}
\put(602.0,839.0){\rule[-0.200pt]{0.400pt}{4.818pt}}
\put(863.0,82.0){\rule[-0.200pt]{0.400pt}{4.818pt}}
\put(863,41){\makebox(0,0){$6$}}
\put(863.0,839.0){\rule[-0.200pt]{0.400pt}{4.818pt}}
\put(1124.0,82.0){\rule[-0.200pt]{0.400pt}{4.818pt}}
\put(1124,41){\makebox(0,0){$8$}}
\put(1124.0,839.0){\rule[-0.200pt]{0.400pt}{4.818pt}}
\put(1385.0,82.0){\rule[-0.200pt]{0.400pt}{4.818pt}}
\put(1385,41){\makebox(0,0){$10$}}
\put(1385.0,839.0){\rule[-0.200pt]{0.400pt}{4.818pt}}
\put(1646.0,82.0){\rule[-0.200pt]{0.400pt}{4.818pt}}
\put(1646,41){\makebox(0,0){$12$}}
\put(1646.0,839.0){\rule[-0.200pt]{0.400pt}{4.818pt}}
\put(1907.0,82.0){\rule[-0.200pt]{0.400pt}{4.818pt}}
\put(1907,41){\makebox(0,0){$14$}}
\put(1907.0,839.0){\rule[-0.200pt]{0.400pt}{4.818pt}}
\put(210.0,82.0){\rule[-0.200pt]{0.400pt}{187.179pt}}
\put(210.0,82.0){\rule[-0.200pt]{440.365pt}{0.400pt}}
\put(2038.0,82.0){\rule[-0.200pt]{0.400pt}{187.179pt}}
\put(210.0,859.0){\rule[-0.200pt]{440.365pt}{0.400pt}}
\put(1110,818){\makebox(0,0)[r]{Quicksort (C library, with compare function)}}
\put(341,116){\makebox(0,0){$+$}}
\put(471,140){\makebox(0,0){$+$}}
\put(602,177){\makebox(0,0){$+$}}
\put(732,226){\makebox(0,0){$+$}}
\put(863,276){\makebox(0,0){$+$}}
\put(993,363){\makebox(0,0){$+$}}
\put(1124,409){\makebox(0,0){$+$}}
\put(1255,481){\makebox(0,0){$+$}}
\put(1385,536){\makebox(0,0){$+$}}
\put(1516,601){\makebox(0,0){$+$}}
\put(1646,667){\makebox(0,0){$+$}}
\put(1777,756){\makebox(0,0){$+$}}
\put(1907,804){\makebox(0,0){$+$}}
\put(1180,818){\makebox(0,0){$+$}}
\put(1110,777){\makebox(0,0)[r]{selection sort (Sedgewick)}}
\put(341,92){\makebox(0,0){$\times$}}
\put(471,102){\makebox(0,0){$\times$}}
\put(602,113){\makebox(0,0){$\times$}}
\put(732,130){\makebox(0,0){$\times$}}
\put(863,150){\makebox(0,0){$\times$}}
\put(993,235){\makebox(0,0){$\times$}}
\put(1124,277){\makebox(0,0){$\times$}}
\put(1255,324){\makebox(0,0){$\times$}}
\put(1385,377){\makebox(0,0){$\times$}}
\put(1516,428){\makebox(0,0){$\times$}}
\put(1646,544){\makebox(0,0){$\times$}}
\put(1777,536){\makebox(0,0){$\times$}}
\put(1907,611){\makebox(0,0){$\times$}}
\put(1180,777){\makebox(0,0){$\times$}}
\put(1110,736){\makebox(0,0)[r]{Quicksort (Sedgewick)}}
\put(341,105){\makebox(0,0){$\ast$}}
\put(471,121){\makebox(0,0){$\ast$}}
\put(602,138){\makebox(0,0){$\ast$}}
\put(732,165){\makebox(0,0){$\ast$}}
\put(863,201){\makebox(0,0){$\ast$}}
\put(993,238){\makebox(0,0){$\ast$}}
\put(1124,279){\makebox(0,0){$\ast$}}
\put(1255,317){\makebox(0,0){$\ast$}}
\put(1385,356){\makebox(0,0){$\ast$}}
\put(1516,391){\makebox(0,0){$\ast$}}
\put(1646,471){\makebox(0,0){$\ast$}}
\put(1777,464){\makebox(0,0){$\ast$}}
\put(1907,501){\makebox(0,0){$\ast$}}
\put(1180,736){\makebox(0,0){$\ast$}}
\put(1110,695){\makebox(0,0)[r]{shell sort (Sedgewick)}}
\put(341,94){\raisebox{-.8pt}{\makebox(0,0){$\Box$}}}
\put(471,101){\raisebox{-.8pt}{\makebox(0,0){$\Box$}}}
\put(602,112){\raisebox{-.8pt}{\makebox(0,0){$\Box$}}}
\put(732,125){\raisebox{-.8pt}{\makebox(0,0){$\Box$}}}
\put(863,169){\raisebox{-.8pt}{\makebox(0,0){$\Box$}}}
\put(993,194){\raisebox{-.8pt}{\makebox(0,0){$\Box$}}}
\put(1124,221){\raisebox{-.8pt}{\makebox(0,0){$\Box$}}}
\put(1255,247){\raisebox{-.8pt}{\makebox(0,0){$\Box$}}}
\put(1385,276){\raisebox{-.8pt}{\makebox(0,0){$\Box$}}}
\put(1516,307){\raisebox{-.8pt}{\makebox(0,0){$\Box$}}}
\put(1646,360){\raisebox{-.8pt}{\makebox(0,0){$\Box$}}}
\put(1777,363){\raisebox{-.8pt}{\makebox(0,0){$\Box$}}}
\put(1907,392){\raisebox{-.8pt}{\makebox(0,0){$\Box$}}}
\put(1180,695){\raisebox{-.8pt}{\makebox(0,0){$\Box$}}}
\put(1110,654){\makebox(0,0)[r]{insertion sort (Sedgewick)}}
\put(341,94){\makebox(0,0){$\blacksquare$}}
\put(471,101){\makebox(0,0){$\blacksquare$}}
\put(602,111){\makebox(0,0){$\blacksquare$}}
\put(732,124){\makebox(0,0){$\blacksquare$}}
\put(863,138){\makebox(0,0){$\blacksquare$}}
\put(993,149){\makebox(0,0){$\blacksquare$}}
\put(1124,184){\makebox(0,0){$\blacksquare$}}
\put(1255,202){\makebox(0,0){$\blacksquare$}}
\put(1385,221){\makebox(0,0){$\blacksquare$}}
\put(1516,246){\makebox(0,0){$\blacksquare$}}
\put(1646,273){\makebox(0,0){$\blacksquare$}}
\put(1777,278){\makebox(0,0){$\blacksquare$}}
\put(1907,300){\makebox(0,0){$\blacksquare$}}
\put(1180,654){\makebox(0,0){$\blacksquare$}}
\put(210.0,82.0){\rule[-0.200pt]{0.400pt}{187.179pt}}
\put(210.0,82.0){\rule[-0.200pt]{440.365pt}{0.400pt}}
\put(2038.0,82.0){\rule[-0.200pt]{0.400pt}{187.179pt}}
\put(210.0,859.0){\rule[-0.200pt]{440.365pt}{0.400pt}}
\end{picture}}
\caption{Comparison of different sorting algorithms for small numbers
  of inputs.}
\label{fig:sorting}
\end{figure}
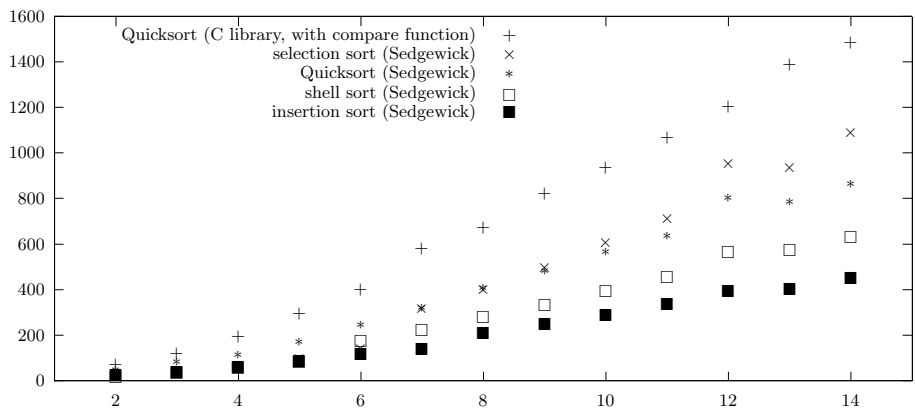

Figure~\ref{fig:sorting}
presents a comparison of the common spectrum of data-dependent sorting
algorithms for small numbers of inputs,
depicting the number of
inputs ($x$-axis) together with the number of cycles
required to sort them ($y$-axis), averaged over $100$ million random
executions.
The upper curve in the figure is obtained from the standard Quicksort
implementation in the C library (which is at some disadvantage, as it
requires a general compare function as an argument). The remaining
curves are derived from applying standard sorting algorithms, as detailed by
Sedgewick~\cite{SedgewickBook}; the code was taken directly from
the book's web page,
\url{http://algs4.cs.princeton.edu/home/}. Insertion sort is the clear winner.

\subsection{Sorting Networks}

A \emph{comparator network} on $n$ channels is a finite sequence
$C=c_1,\ldots,c_k$ of \emph{comparators}, where each comparator
$c_\ell$ is a pair $(i_\ell,j_\ell)$ with $1\leq i_\ell<j_\ell\leq n$.
The \emph{size} of $C$ is the number $k$ of comparators it contains.
Given an input $\vec x\in D^n$, where $D$ is any totally ordered domain,
the \emph{output} of $C$ on $\vec x$ is the sequence $C(\vec x)=\vec x^n$,
where $\vec x^\ell$ is defined inductively as follows:
$\vec x^0=\vec x$, and $\vec x^{\ell}$ is obtained from
$\vec x^{\ell-1}$ by swapping the elements in positions $i_\ell$ and
$j_\ell$, in case $x_{i_\ell}<x_{j_\ell}$.
$C$ is a \emph{sorting network} if
$C(\vec x)$ is sorted for all $C\in D^n$.
It is well known (see e.g.~\cite{Knuth73}) that this property
is independent of the concrete domain $D$.

Comparators may act in parallel.  A comparator network $C$ has
\emph{depth} $d$ if $C$ is the concatenation of $L_1,\ldots,L_d$,
where each $L_i$ is a \emph{layer}: a comparator network with the
property that no two of its comparators act on a common channel.

\begin{figure}[t]
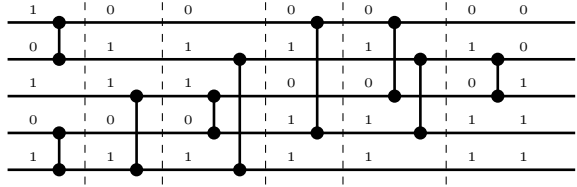

  \centering
  \begin{sortingnetwork}{5}{0.7}
    \nodelabel{\tiny 1,\tiny 0,\tiny 1,\tiny 0,\tiny 1}
    \addcomparator45
    \addcomparator12
    \nextlayer
    \nodelabel{\tiny 1,\tiny 0,\tiny 1,\tiny 1,\tiny 0}
    \addcomparator13
    \nextlayer
    \nodelabel{\tiny 1,\tiny 0,\tiny 1,\tiny 1,\tiny 0}
    \addcomparator23
    \addlayer
    \addcomparator14
    \nextlayer
    \nodelabel{\tiny 1,\tiny 1,\tiny 0,\tiny 1,\tiny 0}
    \addcomparator25
    \nextlayer
    \nodelabel{\tiny 1,\tiny 1,\tiny 0,\tiny 1,\tiny 0}
    \addcomparator35
    \addlayer
    \addcomparator24
    \nextlayer
    \nodelabel{\tiny 1,\tiny 1,\tiny 0,\tiny 1,\tiny 0}
    \addcomparator34
    \addlayer
    \nodelabel{\tiny 1,\tiny 1,\tiny 1,\tiny 0,\tiny 0}
  \end{sortingnetwork}
  \caption{A sorting network on $5$ channels operating on the input $10101$.}
  \label{fig:sort-5}
\end{figure}

Figure~\ref{fig:sort-5} depicts a sorting network on $5$ channels in
the graphical notation we will use throughout this paper.  Comparators
are depicted as vertical lines, and layers are separated by a dashed
line.  The numbers illustrate how the input $10101\in\{0,1\}^5$
propagates through the network.
This network has $6$ layers and
$9$ comparators.

There are two main notions of optimality of sorting networks
in common use: \emph{size}
optimality, where one minimizes the number of comparators used
in the network; and \emph{depth} optimality, where one minimizes
the number of execution steps, taking into account that some
comparators can be executed in parallel.

Given $n$ inputs, finding the minimal size $s_n$ and depth $t_n$ of a
sorting network
is an extremely hard problem that has seen significant progress in recent years.
The table below details
the best currently known bounds.  The values for $n\leq 8$ are already
listed in~\cite{Knuth73}; the values of $t_9$ and $t_{10}$ were proven
exact by Parberry~\cite{Parberry91}, those of $t_{11}$--$t_{16}$ by
Bundala and Z\'avodn\'y~\cite{BZ14}, and $t_{17}$ was recently
computed by Ehlers and M\"uller~\cite{EM15} using results
from~\cite{ourSYNASCpaper,ourLATApaper}.
Finally, the values of $s_9$ and $s_{10}$ were first given in~\cite{ourICTAIpaper}.

\begin{center}
  \small
  \begin{tabular}{c|ccccccccccccccccc}
  \toprule
    $n$ & \,\,1\,\, & \,\,2\,\, & \,\,3\,\, & \,\,4\,\, & \,\,5\,\, &
    \,\,6\,\, & \,\,7\,\, & \,\,8\,\, & \,\,9\,\, & \,10\, & \,11\, &
    \,12\, & \,13\, & \,14\, & \,15\, & \,16\, & \,17\, \\ \midrule
    \multirow2*{$s_n$} & \multirow2*0 & \multirow2*1 & \multirow2*3 &
     \multirow2*5 & \multirow2*9 & \multirow2*{12} & \multirow2*{16} &
     \multirow2*{19} & \multirow2*{25} & \multirow2*{29} &
    35 & 39 & 45 & 51 & 56 & 60 & 73 \\
    &&&&&&&&&&& 33 & 37 & 41 & 45 & 49 & 53 & 58\\ \midrule
    $t_n$ & 0 & 1 & 3 & 3 & 5 & 5 & 6 & 6 & 7 & 7 & 8 & 8 & 9 & 9 & 9 & 9 & 10 \\
    \bottomrule
  \end{tabular}
\end{center}

\begin{figure}
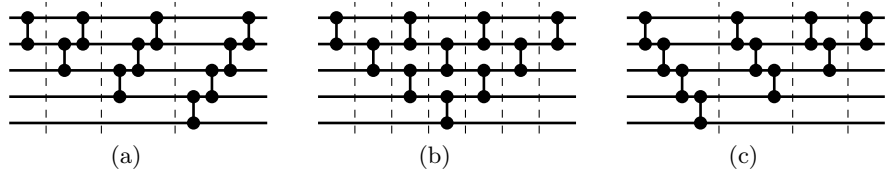

  \centering
  \begin{tabular}{ccc}
  \begin{sortingnetwork}{5}{.5}
    \addcomparator45\nextlayer
    \addcomparator34\addlayer
    \addcomparator45\nextlayer
    \addcomparator23\addlayer
    \addcomparator34\addlayer
    \addcomparator45\nextlayer
    \addcomparator12\addlayer
    \addcomparator23\addlayer
    \addcomparator34\addlayer
    \addcomparator45
  \end{sortingnetwork} &
  \begin{sortingnetwork}{5}{.5}
    \addcomparator45\nextlayer
    \addcomparator34\nextlayer
    \addcomparator45
    \addcomparator23\nextlayer
    \addcomparator34
    \addcomparator12\nextlayer
    \addcomparator45
    \addcomparator23\nextlayer
    \addcomparator34\nextlayer
    \addcomparator45
  \end{sortingnetwork} &
  \begin{sortingnetwork}{5}{.5}
    \addcomparator45\addlayer
    \addcomparator34\addlayer
    \addcomparator23\addlayer
    \addcomparator12\nextlayer
    \addcomparator45\addlayer
    \addcomparator34\addlayer
    \addcomparator23\nextlayer
    \addcomparator45\addlayer
    \addcomparator34\nextlayer
    \addcomparator45
  \end{sortingnetwork} \\
  (a) & (b) & (c)
  \end{tabular}
  \caption{Sorting networks for insertion sort~(a) and bubble-sort~(c) on $5$ inputs,
    dashed lines separating iterations.
    When parallelized, both networks become the same~(b).}
  \label{fig:is-sn}
\end{figure}

Oblivious versions of
classic sorting algorithms can also be
implemented as sorting networks, as described in~\cite{Knuth73}.
Figure~\ref{fig:is-sn}~(a) shows an oblivious version of
insertion-sort.
The vertical dashed lines highlight the $4$
iterations of ``insertion'' required to sort $5$ elements.
Figure~\ref{fig:is-sn}~(b) shows the same network, with comparators
arranged in parallel layers.
Bubble-sort can also be implemented as a sorting network as
illustrated in Figure~\ref{fig:is-sn}~(c), where
the vertical dashed
lines illustrate the $4$ iterations of the classic bubble-sort
algorithm.
When ordered according to layers, this network becomes
identical to the one in Figure~\ref{fig:is-sn}~(b).

\subsection{Modern CPU Architectures}

Modern CPU architectures allow multiple instructions to be performed
in parallel on a single thread. This ability is called
\emph{instruction-level parallelism} (ILP), and is built on three modern
micro-architectural techniques\footnote{For details on these features of modern microarchitectures see e.g.\ \cite{fisher05,silc99}.}:
\begin{itemize}
\item \emph{superscalar instruction pipelines}, i.e., pipelines with the
  ability to hold and execute multiple instructions at the same time
\item \emph{dynamic out-of-order execution}, i.e., dynamic reordering of
  instructions respecting data dependencies
\item \emph{redundant execution units}, i.e., multiple Arithmetic Logic Units per core
\end{itemize}
Together, these features allow execution of instructions in an order
that minimizes data dependencies, so that multiple redundant
execution units can be used at the same time.
This is often termed \emph{implicit} ILP, in contrast to the explicit ILP found in vector operations.

\begin{example}
  Consider
the C expression \texttt{(x+y)*(z+u)}. Assume the variables
\texttt{x}, \texttt{y}, \texttt{z}, and \texttt{u} are loaded in registers
\texttt{eax}, \texttt{ebx}, \texttt{ecx}, and \texttt{edx}. Then the evaluation of the
above expression is compiled to three machine instructions: \texttt{ADD eax,ebx; ADD
ecx,edx; MUL eax,ecx}, with the result in \texttt{ecx}. Here, the first two
instructions are data-independent and can be executed in parallel, while the
last one depends on the results of those, and is executed in another CPU cycle.
\end{example}

Conditional branching instructions are the most expensive instructions
on pipelined CPUs, as they require flushing and refilling the
pipeline. In order to minimize their cost, modern CPU architectures
employ dynamic branch prediction.
By keeping the pipeline filled with the instructions
of the predicted branch, the cost of branching is severely
alleviated. Unfortunately, branch prediction cannot be perfect, and
when the wrong branch is predicted, the pipeline needs to be flushed
and refilled -- an operation taking many CPU cycles.

In order to avoid branching instructions for ``small'' decisions,
e.g., deciding whether to assign a value or not, modern CPU
architectures also feature conditional instructions. Depending on flags
set by e.g.~a comparison, either an assignment of a value of a register will
be performed, or the instruction will be ignored. In both cases,
the pipeline is filled with the subsequent instructions, and the cost of
the operation is smaller than a possible branch prediction
failure.

\begin{example}
  Consider the C statement \texttt{if (x == 42) x = 23;} with variable~\texttt{x} loaded in \texttt{eax}.
  Without conditional move instructions, this
is compiled to code with a conditional branching instruction, i.e.~\texttt{CMP eax,42; JNZ after; MOV eax, 23}, where \texttt{after} is the address of the instruction following the \texttt{MOV} instruction.
Alternatively, using conditional instructions, we obtain \texttt{CMP eax, 42; CMOVZ eax, 23}.
This code not only saves one machine code instruction, but most importantly avoids the huge performance impact of a mispredicted branch.
\end{example}

\section{Quicksort with Sorting Networks for Base Case}
\label{sec:complexity}

The general theme of this paper is to derive, from sorting networks,
optimized code to sort small numbers of inputs, and then to apply this
code as the base case in a Quicksort algorithm.
In this section, we compare precise average case results for the number
of comparisons and swaps performed by a classic Quicksort algorithm
and by a modification that uses sorting networks on subproblems of
size at most $14$. 
We choose $14$ for this analysis, as it is the largest value $n$ for
which we could conveniently measure the number of comparisons and
swaps for all $n!$ permutations. We used the best-known (w.r.t.~size)
sorting networks (optimal for up to $10$ inputs) in order to
obtain the most favorable comparison numbers for sorting networks.  To
this end, we assume the algorithm to act on random permutations of size
$n$, each being the input with equal probability.

Let $C_n$ (resp.\ $S_n$) denote the expected number of comparisons
(resp.\ swaps) performed by classic Quicksort on (random) inputs of
size $n$. Let furthermore $\hat{C}_n$ and $\hat{S}_n$ denote
the corresponding quantities for Quicksort using sorting networks
for inputs smaller than $15$.
It is standard to set up recurrence relations for those quantities
which typically obey a pattern such as:
\[T_n(a,b)=
\begin{cases}
  a\cdot n+b+\frac{1}{n}\sum_{1\le j\le n}T_{j-1}(a,b)+T_{n-j}(a,b) & \mbox{ if $n>M$,}\\
  g(n) & \mbox{ otherwise.}
\end{cases}
\]

Here, $a$ and $b$ have to be chosen properly to reflect the parameter's
(comparisons, swaps) behavior, $M$ determines the maximum subproblem
size for which a different algorithm (insertion sort, sorting
networks) is used, and $g$ accounts for the costs of that algorithm.
In order to analyze classic Quicksort as proposed by Hoare, we have to
choose $a=1$, $b=-1$ (resp.\ $a=\frac{1}{6}$, $b=\frac{2}{3}$) for
comparisons (resp.\ swaps), together with $M=0$ and $g(0)=0$.  For the
analysis of our proposed modification using sorting networks for
subproblems of small sizes, we set $M=14$ together with the values for
$g$ as given in Table~\ref{tab_functiong}.  Using standard algebraic
manipulations, it is possible to solve this recurrence explicitly to
obtain a formula for $T_n(a,b)$ in terms of $n$, $M$, $a$ and
$b$. Defining $t_n=a\cdot n+b$ and $\nabla t_n=t_n-t_{n-1}$, one finds
(see \cite{sedgewick} for details) that, for $n>M$,
$$T_n(a,b) = 2(n+1)\sum_{M+2\le k\le n}\frac{\nabla t_k}{k+1} +
\frac{n+1}{M+2}\left(t_{M+1} + T_{M+1}(a,b)\right) - t_n\,.$$

\begin{table}[t]
\caption{Average number of comparisons and swaps when executing
  optimal sorting networks with at most $M=14$ inputs.}
  \label{tab_functiong}
\centering
\begin{tabular}{c|cccccccccccccc}
\toprule
$n$         & $1$ & $2$ & $3$ & $4$ & $5$ & $6$ & $7$ & $8$ & $9$ & $10$ & $11$ & $12$ & $13$ & $14$ \\
\midrule
comparisons & $0$ & $1$ & $3$ & $5$ & $9$ & $12$ & $16$ & $19$ & $25$ & $29$ & $35$ & $39$ & $45$ & $51$ \\
\midrule
swaps       & \,$0.0$\,&\,$0.5$\,&\,$1.5$\,&\,$2.7$\,&\,$4.8$\,&\,$6.6$\,&\,$8.6$\,&\,$10.6$\,&\,$13.0$\,&\,$11.1$\,&\,$19.4$\,&\,$22.4$\,&\,$20.0$\,&\,$26.5$ \\
\bottomrule
\end{tabular}

\end{table}

Computing the closed form expressions for
$\sum_{M+2\le k\le n}\frac{\nabla t_k}{k+1}$ for the different choices of $t_n$, we
finally get
\begin{align*}
C_n&=2n\ln(n+1)-2.84557 n+o(n) &
S_n&=\frac{1}{3}n\ln(n+1)+0.359072 n+o(n) \\
\hat{C}_n&=2n\ln(n+1)-2.44869 n+o(n) &
\,\,\,\hat{S}_n&=\frac{1}{3}n\ln(n+1)+ 0.524887 n+o(n)
\end{align*}
We see that, when increasing $n$, both parameters get worse by
our modification of classic Quicksort. Even for small $n$ and optimal
size sorting networks, there is no advantage w.r.t.\ the numbers of
comparisons or swaps. In conclusion, we cannot hope to get a faster
sorting algorithm simply by switching to sorting networks for small
subproblems -- at least not on grounds of our theoretical
investigations. And, by transitivity, replacing insertion sort by
sorting networks in the base case should result in an even worse
behavior w.r.t.\ both parameters.

\section{Unrolling the Base Case}
\label{sec:unrolling}
In this section, we show how to unroll an
implementation of insertion sort, step by step, until we finally
obtain code equivalent to a sorting network. We take the
basic insertion sort code from Sedgewick \cite{SedgewickBook}, and, for
illustration, assume that the fixed number of inputs is $n=5$. We
experimented also with optimized variants (e.g.~making use of
sentinels to avoid the \texttt{j>0} check), but did not find any of
them to be faster for small inputs given a modern C compiler.

\begin{lstlisting}
#define SWAP(x,y) {int tmp = a[x]; a[x] = a[y]; a[y] = tmp;}
static inline void sort5(int *a, int n) {
  n=5
  for (int i = 1; i < n; i++)
    for (int j = i; j > 0 && a[j] < a[j-1]; j--)
      SWAP(j-1, j)
}
\end{lstlisting}
Applying partial evaluation and (outer) loop unrolling
results in:
\begin{lstlisting}
static inline void sort5_unrolled(int *a) {
  for (int j = 1; j > 0 && a[j] < a[j-1]; j--)
    SWAP(j-1, j)
  for (int j = 2; j > 0 && a[j] < a[j-1]; j--)
    SWAP(j-1, j)
  for (int j = 3; j > 0 && a[j] < a[j-1]; j--)
    SWAP(j-1, j)
  for (int j = 4; j > 0 && a[j] < a[j-1]; j--)
    SWAP(j-1, j)
}
\end{lstlisting}
The condition in the inner loop is data-dependent, hence no sound and
complete program transformation can be applied to unroll them. To
address this, we move the data-dependent part of the loop condition to
the statement in the body of the loop, while always iterating the
variable \texttt{j} down to $1$.
\begin{lstlisting}
static inline void sort5_oblivious(int *a) {
  for (int j = 1; j > 0; j--)
    if (a[j] < a[j-1]) SWAP(j-1, j)
  for (int j = 2; j > 0; j--)
    if (a[j] < a[j-1]) SWAP(j-1, j)
  for (int j = 3; j > 0; j--)
    if (a[j] < a[j-1]) SWAP(j-1, j)
  for (int j = 4; j > 0; j--)
    if (a[j] < a[j-1]) SWAP(j-1, j)
}
\end{lstlisting}
Now we can now apply (inner) loop unrolling  and obtain:
\begin{lstlisting}
static inline void sort5_oblivous_unrolled(int *a) {
  if (a[1] < a[0]) SWAP(0, 1)
  if (a[2] < a[1]) SWAP(1, 2)
  if (a[1] < a[0]) SWAP(0, 1)
  if (a[3] < a[2]) SWAP(2, 3)
  if (a[2] < a[1]) SWAP(1, 2)
  if (a[1] < a[0]) SWAP(0, 1)
  if (a[4] < a[3]) SWAP(3, 4)
  if (a[3] < a[2]) SWAP(2, 3)
  if (a[2] < a[1]) SWAP(1, 2)
  if (a[1] < a[0]) SWAP(0, 1)
}
\end{lstlisting}
All the statements in the body of
\texttt{sort5\_oblivous\_unrolled} are now conditional swaps. For
readability, we move the condition into the macro. \texttt{\color{magenta}COMP}s on
the same line indicate that they originate from the same iteration of
insertion sort:
\begin{lstlisting}
#define COMP(x,y) { if (a[y] < a[x]) SWAP(x,y) }
static inline void sort5_fig3a(int *a) {
  COMP(0, 1)
  COMP(1, 2)  COMP(0, 1)
  COMP(2, 3)  COMP(1, 2)  COMP(0, 1)
  COMP(3, 4)  COMP(2, 3)  COMP(1, 2)  COMP(0, 1)
}
\end{lstlisting}
This sequence is equivalent to the sorting network in
Figure~\ref{fig:is-sn}~(a). Thus, we can apply the reordering of
comparators that resulted in Figure~\ref{fig:is-sn}~(b) to obtain the
following implementation, where we reduce the number of layers to $7$
(here, \texttt{\color{magenta}COMP}s on the same line indicate a layer in the
sorting network):
\begin{lstlisting}
static inline void sort5_fig3b(int *a) {
  COMP(0, 1)
  COMP(1, 2)
  COMP(0, 1)  COMP(2, 3)
  COMP(1, 2)  COMP(3, 4)
  COMP(0, 1)  COMP(2, 3)
  COMP(1, 2)
  COMP(0, 1)
}
\end{lstlisting}

\begin{figure}[t]
\centering
\resizebox{\textwidth}{!}{
\setlength{\unitlength}{0.240900pt}
\ifx\plotpoint\undefined\newsavebox{\plotpoint}\fi
\sbox{\plotpoint}{\rule[-0.200pt]{0.400pt}{0.400pt}}%
\begin{picture}(1930,850)(130,50)
\sbox{\plotpoint}{\rule[-0.200pt]{0.400pt}{0.400pt}}%
\put(170.0,82.0){\rule[-0.200pt]{4.818pt}{0.400pt}}
\put(150,82){\makebox(0,0)[r]{$0$}}
\put(2018.0,82.0){\rule[-0.200pt]{4.818pt}{0.400pt}}
\put(170.0,193.0){\rule[-0.200pt]{4.818pt}{0.400pt}}
\put(150,193){\makebox(0,0)[r]{$100$}}
\put(2018.0,193.0){\rule[-0.200pt]{4.818pt}{0.400pt}}
\put(170.0,304.0){\rule[-0.200pt]{4.818pt}{0.400pt}}
\put(150,304){\makebox(0,0)[r]{$200$}}
\put(2018.0,304.0){\rule[-0.200pt]{4.818pt}{0.400pt}}
\put(170.0,415.0){\rule[-0.200pt]{4.818pt}{0.400pt}}
\put(150,415){\makebox(0,0)[r]{$300$}}
\put(2018.0,415.0){\rule[-0.200pt]{4.818pt}{0.400pt}}
\put(170.0,526.0){\rule[-0.200pt]{4.818pt}{0.400pt}}
\put(150,526){\makebox(0,0)[r]{$400$}}
\put(2018.0,526.0){\rule[-0.200pt]{4.818pt}{0.400pt}}
\put(170.0,637.0){\rule[-0.200pt]{4.818pt}{0.400pt}}
\put(150,637){\makebox(0,0)[r]{$500$}}
\put(2018.0,637.0){\rule[-0.200pt]{4.818pt}{0.400pt}}
\put(170.0,748.0){\rule[-0.200pt]{4.818pt}{0.400pt}}
\put(150,748){\makebox(0,0)[r]{$600$}}
\put(2018.0,748.0){\rule[-0.200pt]{4.818pt}{0.400pt}}
\put(170.0,859.0){\rule[-0.200pt]{4.818pt}{0.400pt}}
\put(150,859){\makebox(0,0)[r]{$700$}}
\put(2018.0,859.0){\rule[-0.200pt]{4.818pt}{0.400pt}}
\put(303.0,82.0){\rule[-0.200pt]{0.400pt}{4.818pt}}
\put(303,41){\makebox(0,0){$2$}}
\put(303.0,839.0){\rule[-0.200pt]{0.400pt}{4.818pt}}
\put(570.0,82.0){\rule[-0.200pt]{0.400pt}{4.818pt}}
\put(570,41){\makebox(0,0){$4$}}
\put(570.0,839.0){\rule[-0.200pt]{0.400pt}{4.818pt}}
\put(837.0,82.0){\rule[-0.200pt]{0.400pt}{4.818pt}}
\put(837,41){\makebox(0,0){$6$}}
\put(837.0,839.0){\rule[-0.200pt]{0.400pt}{4.818pt}}
\put(1104.0,82.0){\rule[-0.200pt]{0.400pt}{4.818pt}}
\put(1104,41){\makebox(0,0){$8$}}
\put(1104.0,839.0){\rule[-0.200pt]{0.400pt}{4.818pt}}
\put(1371.0,82.0){\rule[-0.200pt]{0.400pt}{4.818pt}}
\put(1371,41){\makebox(0,0){$10$}}
\put(1371.0,839.0){\rule[-0.200pt]{0.400pt}{4.818pt}}
\put(1638.0,82.0){\rule[-0.200pt]{0.400pt}{4.818pt}}
\put(1638,41){\makebox(0,0){$12$}}
\put(1638.0,839.0){\rule[-0.200pt]{0.400pt}{4.818pt}}
\put(1905.0,82.0){\rule[-0.200pt]{0.400pt}{4.818pt}}
\put(1905,41){\makebox(0,0){$14$}}
\put(1905.0,839.0){\rule[-0.200pt]{0.400pt}{4.818pt}}
\put(170.0,82.0){\rule[-0.200pt]{0.400pt}{187.179pt}}
\put(170.0,82.0){\rule[-0.200pt]{450.001pt}{0.400pt}}
\put(2038.0,82.0){\rule[-0.200pt]{0.400pt}{187.179pt}}
\put(170.0,859.0){\rule[-0.200pt]{450.001pt}{0.400pt}}
\put(1090,818){\makebox(0,0)[r]{insertion sort}}
\put(303,109){\makebox(0,0){$+$}}
\put(437,126){\makebox(0,0){$+$}}
\put(570,149){\makebox(0,0){$+$}}
\put(704,177){\makebox(0,0){$+$}}
\put(837,211){\makebox(0,0){$+$}}
\put(971,235){\makebox(0,0){$+$}}
\put(1104,316){\makebox(0,0){$+$}}
\put(1237,357){\makebox(0,0){$+$}}
\put(1371,400){\makebox(0,0){$+$}}
\put(1504,456){\makebox(0,0){$+$}}
\put(1638,519){\makebox(0,0){$+$}}
\put(1771,530){\makebox(0,0){$+$}}
\put(1905,581){\makebox(0,0){$+$}}
\put(1160,818){\makebox(0,0){$+$}}
\put(1090,777){\makebox(0,0)[r]{unrolled insertion sort}}
\put(303,109){\makebox(0,0){$\times$}}
\put(437,127){\makebox(0,0){$\times$}}
\put(570,150){\makebox(0,0){$\times$}}
\put(704,177){\makebox(0,0){$\times$}}
\put(837,215){\makebox(0,0){$\times$}}
\put(971,248){\makebox(0,0){$\times$}}
\put(1104,278){\makebox(0,0){$\times$}}
\put(1237,316){\makebox(0,0){$\times$}}
\put(1371,355){\makebox(0,0){$\times$}}
\put(1504,401){\makebox(0,0){$\times$}}
\put(1638,427){\makebox(0,0){$\times$}}
\put(1771,483){\makebox(0,0){$\times$}}
\put(1905,517){\makebox(0,0){$\times$}}
\put(1160,777){\makebox(0,0){$\times$}}
\put(1090,736){\makebox(0,0)[r]{insertion sorting network}}
\put(303,109){\makebox(0,0){$\ast$}}
\put(437,123){\makebox(0,0){$\ast$}}
\put(570,147){\makebox(0,0){$\ast$}}
\put(704,175){\makebox(0,0){$\ast$}}
\put(837,212){\makebox(0,0){$\ast$}}
\put(971,247){\makebox(0,0){$\ast$}}
\put(1104,289){\makebox(0,0){$\ast$}}
\put(1237,328){\makebox(0,0){$\ast$}}
\put(1371,405){\makebox(0,0){$\ast$}}
\put(1504,507){\makebox(0,0){$\ast$}}
\put(1638,584){\makebox(0,0){$\ast$}}
\put(1771,663){\makebox(0,0){$\ast$}}
\put(1905,773){\makebox(0,0){$\ast$}}
\put(1160,736){\makebox(0,0){$\ast$}}
\put(1090,695){\makebox(0,0)[r]{compressed insertion sorting network}}
\put(303,108){\raisebox{-.8pt}{\makebox(0,0){$\Box$}}}
\put(437,122){\raisebox{-.8pt}{\makebox(0,0){$\Box$}}}
\put(570,145){\raisebox{-.8pt}{\makebox(0,0){$\Box$}}}
\put(704,172){\raisebox{-.8pt}{\makebox(0,0){$\Box$}}}
\put(837,209){\raisebox{-.8pt}{\makebox(0,0){$\Box$}}}
\put(971,244){\raisebox{-.8pt}{\makebox(0,0){$\Box$}}}
\put(1104,286){\raisebox{-.8pt}{\makebox(0,0){$\Box$}}}
\put(1237,331){\raisebox{-.8pt}{\makebox(0,0){$\Box$}}}
\put(1371,384){\raisebox{-.8pt}{\makebox(0,0){$\Box$}}}
\put(1504,438){\raisebox{-.8pt}{\makebox(0,0){$\Box$}}}
\put(1638,487){\raisebox{-.8pt}{\makebox(0,0){$\Box$}}}
\put(1771,540){\raisebox{-.8pt}{\makebox(0,0){$\Box$}}}
\put(1905,632){\raisebox{-.8pt}{\makebox(0,0){$\Box$}}}
\put(1160,695){\raisebox{-.8pt}{\makebox(0,0){$\Box$}}}
\put(1090,654){\makebox(0,0)[r]{optimal sorting network}}
\put(303,109){\makebox(0,0){$\blacksquare$}}
\put(437,122){\makebox(0,0){$\blacksquare$}}
\put(570,141){\makebox(0,0){$\blacksquare$}}
\put(704,168){\makebox(0,0){$\blacksquare$}}
\put(837,201){\makebox(0,0){$\blacksquare$}}
\put(971,242){\makebox(0,0){$\blacksquare$}}
\put(1104,265){\makebox(0,0){$\blacksquare$}}
\put(1237,296){\makebox(0,0){$\blacksquare$}}
\put(1371,357){\makebox(0,0){$\blacksquare$}}
\put(1504,395){\makebox(0,0){$\blacksquare$}}
\put(1638,421){\makebox(0,0){$\blacksquare$}}
\put(1771,471){\makebox(0,0){$\blacksquare$}}
\put(1905,523){\makebox(0,0){$\blacksquare$}}
\put(1160,654){\makebox(0,0){$\blacksquare$}}
\put(170.0,82.0){\rule[-0.200pt]{0.400pt}{187.179pt}}
\put(170.0,82.0){\rule[-0.200pt]{450.001pt}{0.400pt}}
\put(2038.0,82.0){\rule[-0.200pt]{0.400pt}{187.179pt}}
\put(170.0,859.0){\rule[-0.200pt]{450.001pt}{0.400pt}}
\end{picture}}
\caption{Comparison of insertion sort with (unrolled) comparator based
  code for small numbers of inputs.}
\label{fig:insertion}
\end{figure}
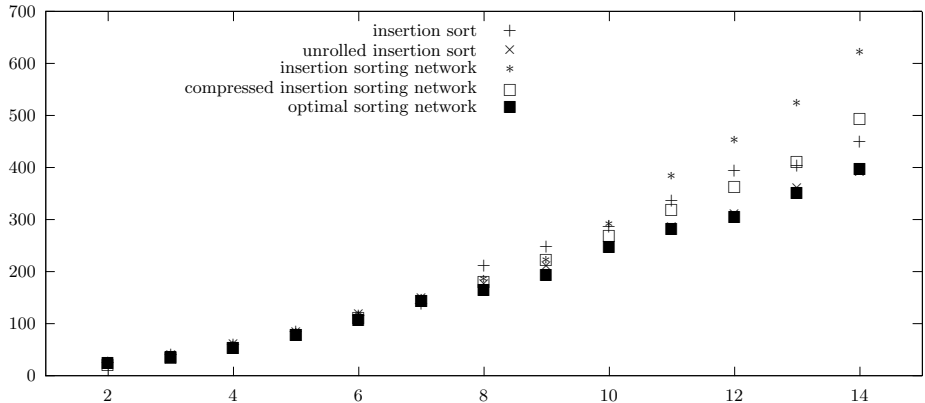

Figure~\ref{fig:insertion} presents a comparison of a standard
insertion sort (code from \cite{SedgewickBook}) with the several
optimized versions, depicting the number of inputs ($x$-axis)
together with the number of cycles required to sort them
($y$-axis), averaged over $100$ million random executions.
The curve labeled ``insertion sort'' portrays the same data as the
corresponding curve in Figure~\ref{fig:sorting}. The curve labeled
``unrolled insertion sort'' corresponds to the unrolled version of
insertion sort (in the style of function \texttt{sort5\_unrolled}).
The other three curves correspond to code derived from different types
of sorting networks: the ``insertion sorting network'' from
Figure~\ref{fig:is-sn}~(a) and function \texttt{sort5\_fig3a};
the ``compressed insertion sorting network'' from
Figure~\ref{fig:is-sn}~(b) and function \texttt{sort5\_fig3b}; and
the ``optimal sorting network'', corresponding to the use of a best
(smallest) known sorting network.

From the figure, it is clear that standard sorting network optimizations such
as reordering of independent comparators \cite{Knuth73} give a slight
performance boost.
But there is another clear message: even going beyond
standard program transformations by breaking data-dependence and
obtaining a sequence of conditional swaps (i.e., a sorting network),
we do not manage to make any significant improvements of the
performance of sorting implementations for small numbers of
inputs. Furthermore, even when using size-optimal sorting networks, we
obtain no real benefit over compiler-optimized insertion sort. This is
in line with the theoretical results on average case complexity
discussed in the previous section.

\section{Implementing Sorting Networks Efficiently}
\label{sec:implementing}

The results in the previous two sections explained the rather
discouraging results obtained by a naive attempt to use sorting networks
as the base case of a divide-and-conquer sorting algorithm: they are
simply not faster than e.g.~insertion sort -- at least when
implemented naively.
In this section we show how to exploit two main properties of
sorting networks, together with features of modern CPU architectures,
and obtain speed-ups of a factor higher than $3$ compared to
unrolled insertion sort.

We first observe that, as sorting networks are data-oblivious, the order of
comparisons is fully determined at compile time, i.e.,
their implementation is
free of any control-flow branching.  Unfortunately,
the naive implementation of each comparator involves branching to
decide whether to perform a swap. The path taken depends
entirely on the specific inputs to be sorted, and as such branch
prediction necessarily does not perform very well.

Luckily, we can also implement comparators without branching.  To
this end, we use a conditional assignment (defined by the macro
\texttt{\color{magenta}COND} below), which can be compiled to the
conditonal move (\texttt{CMOV}) instruction available on modern CPU
architectures. This approach proved to be very fruitful. For
illustration, from the optimal-size sorting network for $5$ inputs
portrayed in Figure~\ref{fig:sort-5},
we synthesize the following C function \texttt{sort5\_best}, where each
row in the code corresponds to a layer in the sorting network:
\begin{lstlisting}
#define COND(c,x,y) { x = (c) ? y : x; }
#define COMP(x,y) { int ax = a[x]; COND(a[y]<ax,a[x],a[y]); \COND(a[y]<ax,a[y],ax  ); }

static inline void sort5_best(int *a) {
  COMP(0, 1)  COMP(3, 4)
  COMP(2, 4)
  COMP(2, 3)  COMP(1, 4)
  COMP(0, 3)
  COMP(0, 2)  COMP(1, 3)
  COMP(1, 2)
}
\end{lstlisting}
The comparator macro that compares and conditionally swaps the values
at indices \texttt{\color{magenta}x} and \texttt{\color{magenta}y}
works as follows:
\begin{enumerate}
\item Keep a copy of the value at index \texttt{\color{magenta}x}.
\item Compare (once) the value at index \texttt{\color{magenta}y} with
  the stored value from \texttt{\color{magenta}x}.
\item If the value was greater, copy the value at index
  \texttt{\color{magenta}y} to index
  \texttt{\color{magenta}x}. Otherwise, do nothing.
\item If the value was greater, write the old copied value from
  \texttt{\color{magenta}x} to index
  \texttt{\color{magenta}y}. Otherwise, do nothing.
\end{enumerate}
Correctness follows directly by case analysis.
If the value at index
\texttt{\color{magenta}y} was not greater than the value at index
\texttt{\color{magenta}x}, the two conditional assignments do not
change anything, and all we did was an unnecessary copy of the valued
at index \texttt{\color{magenta}x}. If the value at index
\texttt{\color{magenta}y} was greater than the value at index
\texttt{\color{magenta}y}, we essentially perform a classic swap using
\texttt{\color{magenta}ax} as the temporary variable.

Given a sufficient optimization level (\texttt{-O2} and above), the
above code is compiled by the LLVM (or GNU) C compiler to use two conditional
move (\texttt{CMOV}) instructions, resulting in a totally branching free code
for \texttt{sort5\_best}. As can be expected, the other two
instructions are a move (\texttt{MOV}) and a compare (\texttt{CMP}) instruction. In
other words, each comparator is implemented by exactly four
non-branching machine code instructions.

Alternatively, we could implement the comparator applying the folklore
idea of swapping values using \texttt{XOR}s to eliminate one conditional
assignment:\footnote{See
  \url{https://graphics.stanford.edu/\~{}seander/bithacks.html\#SwappingValuesXOR}}
\begin{lstlisting}
#define COND(c,x,y) { x = (c) ? y : x; }
#define COMP(x,y) { int ax = a[x]; COND(a[y]<ax,a[x],a[y]); \a[y] ^= ax ^ a[x]; }
\end{lstlisting}
This alternative comparator performs a conditional swap as
follows:
\begin{enumerate}
\item Keep a copy of the value at index \texttt{\color{magenta}x}.
\item If the value at index \texttt{\color{magenta}y} is greater than
  the value at index \texttt{\color{magenta}x}, copy the value at
  index \texttt{\color{magenta}y} to index \texttt{\color{magenta}x}.
\item Bitwise XOR the value at index \texttt{\color{magenta}y} with
  the copied old and the new value at index \texttt{\color{magenta}x}.
\end{enumerate}
Step 3 works because, if the condition holds, then
\texttt{\color{magenta}ax} and the value at index
\texttt{\color{magenta}x} cancel out, leaving the value at
\texttt{\color{magenta}y} unchanged, while otherwise the value at
\texttt{\color{magenta}y} and \texttt{\color{magenta}ax} cancel out,
effectively assigning the original value from index
\texttt{\color{magenta}x} to index \texttt{\color{magenta}y}.

We also implemented this variant, and observed that it compiles down to
five instructions (\texttt{MOV}, \texttt{CMP}, \texttt{CMOV}, and two \texttt{XORs}). We benchmarked the
two variants and observed that they are indistinguishable in practice,
with differences well within the margin of measurement error.  Thus,
we decided to continue with this second version, as the \texttt{XOR} instructions are
more ``basic'' and can therefore be expected to behave better w.r.t.\
e.g. instruction level parallelism.

A third approach would be to define branching-free minimum and
maximum operations,\footnote{See
  \url{https://graphics.stanford.edu/\~{}seander/bithacks.html\#IntegerMinOrMax}}
and use them to assign the minimum to the upper channel and the
maximum to the lower channel of the comparator. We tested this
approach, but found that it did not compile to branching-free
code. Even if it did, the number of instructions involved would be
rather large, eliminating any chance of competing with the two
previous variants.

The reader might wonder whether a different
\texttt{\color{magenta}SWAP} macro could similarly speed up the
working of standard insertion sort. The answer is a clear no, as
the standard swapping operation is implemented by only three
operations. Tricks like using \texttt{XOR}s only increase the number of
instructions to execute, while not reducing branching
in the code. We implemented and benchmarked several
alternative \texttt{\color{magenta}SWAP} macros, finding only
detrimental effects on measured performance.

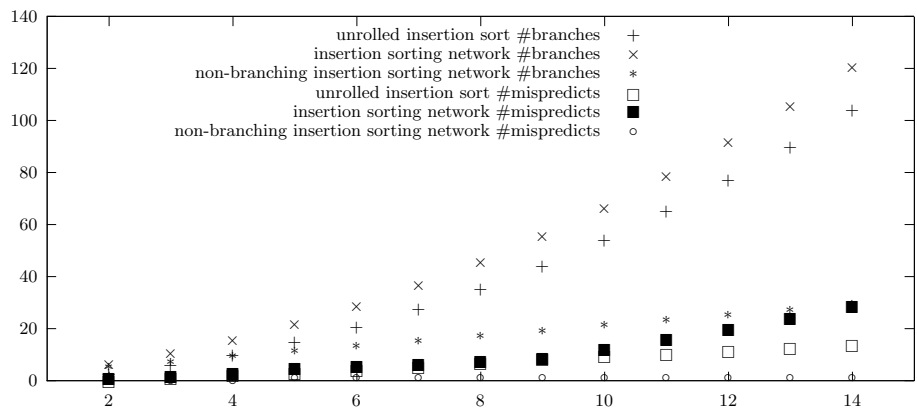
\begin{figure}[t]
\centering
\resizebox{\textwidth}{!}{
\setlength{\unitlength}{0.240900pt}
\ifx\plotpoint\undefined\newsavebox{\plotpoint}\fi
\sbox{\plotpoint}{\rule[-0.200pt]{0.400pt}{0.400pt}}%
\begin{picture}(1930,850)(130,50)
\sbox{\plotpoint}{\rule[-0.200pt]{0.400pt}{0.400pt}}%
\put(190.0,82.0){\rule[-0.200pt]{4.818pt}{0.400pt}}
\put(170,82){\makebox(0,0)[r]{$0$}}
\put(2018.0,82.0){\rule[-0.200pt]{4.818pt}{0.400pt}}
\put(190.0,193.0){\rule[-0.200pt]{4.818pt}{0.400pt}}
\put(170,193){\makebox(0,0)[r]{$20$}}
\put(2018.0,193.0){\rule[-0.200pt]{4.818pt}{0.400pt}}
\put(190.0,304.0){\rule[-0.200pt]{4.818pt}{0.400pt}}
\put(170,304){\makebox(0,0)[r]{$40$}}
\put(2018.0,304.0){\rule[-0.200pt]{4.818pt}{0.400pt}}
\put(190.0,415.0){\rule[-0.200pt]{4.818pt}{0.400pt}}
\put(170,415){\makebox(0,0)[r]{$60$}}
\put(2018.0,415.0){\rule[-0.200pt]{4.818pt}{0.400pt}}
\put(190.0,526.0){\rule[-0.200pt]{4.818pt}{0.400pt}}
\put(170,526){\makebox(0,0)[r]{$80$}}
\put(2018.0,526.0){\rule[-0.200pt]{4.818pt}{0.400pt}}
\put(190.0,637.0){\rule[-0.200pt]{4.818pt}{0.400pt}}
\put(170,637){\makebox(0,0)[r]{$100$}}
\put(2018.0,637.0){\rule[-0.200pt]{4.818pt}{0.400pt}}
\put(190.0,748.0){\rule[-0.200pt]{4.818pt}{0.400pt}}
\put(170,748){\makebox(0,0)[r]{$120$}}
\put(2018.0,748.0){\rule[-0.200pt]{4.818pt}{0.400pt}}
\put(190.0,859.0){\rule[-0.200pt]{4.818pt}{0.400pt}}
\put(170,859){\makebox(0,0)[r]{$140$}}
\put(2018.0,859.0){\rule[-0.200pt]{4.818pt}{0.400pt}}
\put(322.0,82.0){\rule[-0.200pt]{0.400pt}{4.818pt}}
\put(322,41){\makebox(0,0){$2$}}
\put(322.0,839.0){\rule[-0.200pt]{0.400pt}{4.818pt}}
\put(586.0,82.0){\rule[-0.200pt]{0.400pt}{4.818pt}}
\put(586,41){\makebox(0,0){$4$}}
\put(586.0,839.0){\rule[-0.200pt]{0.400pt}{4.818pt}}
\put(850.0,82.0){\rule[-0.200pt]{0.400pt}{4.818pt}}
\put(850,41){\makebox(0,0){$6$}}
\put(850.0,839.0){\rule[-0.200pt]{0.400pt}{4.818pt}}
\put(1114.0,82.0){\rule[-0.200pt]{0.400pt}{4.818pt}}
\put(1114,41){\makebox(0,0){$8$}}
\put(1114.0,839.0){\rule[-0.200pt]{0.400pt}{4.818pt}}
\put(1378.0,82.0){\rule[-0.200pt]{0.400pt}{4.818pt}}
\put(1378,41){\makebox(0,0){$10$}}
\put(1378.0,839.0){\rule[-0.200pt]{0.400pt}{4.818pt}}
\put(1642.0,82.0){\rule[-0.200pt]{0.400pt}{4.818pt}}
\put(1642,41){\makebox(0,0){$12$}}
\put(1642.0,839.0){\rule[-0.200pt]{0.400pt}{4.818pt}}
\put(1906.0,82.0){\rule[-0.200pt]{0.400pt}{4.818pt}}
\put(1906,41){\makebox(0,0){$14$}}
\put(1906.0,839.0){\rule[-0.200pt]{0.400pt}{4.818pt}}
\put(190.0,82.0){\rule[-0.200pt]{0.400pt}{187.179pt}}
\put(190.0,82.0){\rule[-0.200pt]{445.183pt}{0.400pt}}
\put(2038.0,82.0){\rule[-0.200pt]{0.400pt}{187.179pt}}
\put(190.0,859.0){\rule[-0.200pt]{445.183pt}{0.400pt}}
\put(1370,818){\makebox(0,0)[r]{unrolled insertion sort \#branches}}
\put(322,95){\makebox(0,0){$+$}}
\put(454,115){\makebox(0,0){$+$}}
\put(586,136){\makebox(0,0){$+$}}
\put(718,163){\makebox(0,0){$+$}}
\put(850,195){\makebox(0,0){$+$}}
\put(982,233){\makebox(0,0){$+$}}
\put(1114,277){\makebox(0,0){$+$}}
\put(1246,326){\makebox(0,0){$+$}}
\put(1378,381){\makebox(0,0){$+$}}
\put(1510,442){\makebox(0,0){$+$}}
\put(1642,508){\makebox(0,0){$+$}}
\put(1774,580){\makebox(0,0){$+$}}
\put(1906,657){\makebox(0,0){$+$}}
\put(1440,818){\makebox(0,0){$+$}}
\put(1370,777){\makebox(0,0)[r]{insertion sorting network \#branches}}
\put(322,117){\makebox(0,0){$\times$}}
\put(454,139){\makebox(0,0){$\times$}}
\put(586,167){\makebox(0,0){$\times$}}
\put(718,201){\makebox(0,0){$\times$}}
\put(850,239){\makebox(0,0){$\times$}}
\put(982,284){\makebox(0,0){$\times$}}
\put(1114,334){\makebox(0,0){$\times$}}
\put(1246,389){\makebox(0,0){$\times$}}
\put(1378,450){\makebox(0,0){$\times$}}
\put(1510,517){\makebox(0,0){$\times$}}
\put(1642,589){\makebox(0,0){$\times$}}
\put(1774,667){\makebox(0,0){$\times$}}
\put(1906,750){\makebox(0,0){$\times$}}
\put(1440,777){\makebox(0,0){$\times$}}
\put(1370,736){\makebox(0,0)[r]{non-branching insertion sorting network \#branches}}
\put(322,112){\makebox(0,0){$\ast$}}
\put(454,123){\makebox(0,0){$\ast$}}
\put(586,134){\makebox(0,0){$\ast$}}
\put(718,145){\makebox(0,0){$\ast$}}
\put(850,156){\makebox(0,0){$\ast$}}
\put(982,167){\makebox(0,0){$\ast$}}
\put(1114,178){\makebox(0,0){$\ast$}}
\put(1246,189){\makebox(0,0){$\ast$}}
\put(1378,201){\makebox(0,0){$\ast$}}
\put(1510,212){\makebox(0,0){$\ast$}}
\put(1642,223){\makebox(0,0){$\ast$}}
\put(1774,234){\makebox(0,0){$\ast$}}
\put(1906,245){\makebox(0,0){$\ast$}}
\put(1440,736){\makebox(0,0){$\ast$}}
\put(1370,695){\makebox(0,0)[r]{unrolled insertion sort \#mispredicts}}
\put(322,82){\raisebox{-.8pt}{\makebox(0,0){$\Box$}}}
\put(454,89){\raisebox{-.8pt}{\makebox(0,0){$\Box$}}}
\put(586,95){\raisebox{-.8pt}{\makebox(0,0){$\Box$}}}
\put(718,100){\raisebox{-.8pt}{\makebox(0,0){$\Box$}}}
\put(850,106){\raisebox{-.8pt}{\makebox(0,0){$\Box$}}}
\put(982,113){\raisebox{-.8pt}{\makebox(0,0){$\Box$}}}
\put(1114,121){\raisebox{-.8pt}{\makebox(0,0){$\Box$}}}
\put(1246,129){\raisebox{-.8pt}{\makebox(0,0){$\Box$}}}
\put(1378,135){\raisebox{-.8pt}{\makebox(0,0){$\Box$}}}
\put(1510,141){\raisebox{-.8pt}{\makebox(0,0){$\Box$}}}
\put(1642,147){\raisebox{-.8pt}{\makebox(0,0){$\Box$}}}
\put(1774,153){\raisebox{-.8pt}{\makebox(0,0){$\Box$}}}
\put(1906,159){\raisebox{-.8pt}{\makebox(0,0){$\Box$}}}
\put(1440,695){\raisebox{-.8pt}{\makebox(0,0){$\Box$}}}
\put(1370,654){\makebox(0,0)[r]{insertion sorting network \#mispredicts}}
\put(322,85){\makebox(0,0){$\blacksquare$}}
\put(454,90){\makebox(0,0){$\blacksquare$}}
\put(586,97){\makebox(0,0){$\blacksquare$}}
\put(718,106){\makebox(0,0){$\blacksquare$}}
\put(850,110){\makebox(0,0){$\blacksquare$}}
\put(982,116){\makebox(0,0){$\blacksquare$}}
\put(1114,122){\makebox(0,0){$\blacksquare$}}
\put(1246,129){\makebox(0,0){$\blacksquare$}}
\put(1378,148){\makebox(0,0){$\blacksquare$}}
\put(1510,168){\makebox(0,0){$\blacksquare$}}
\put(1642,191){\makebox(0,0){$\blacksquare$}}
\put(1774,214){\makebox(0,0){$\blacksquare$}}
\put(1906,240){\makebox(0,0){$\blacksquare$}}
\put(1440,654){\makebox(0,0){$\blacksquare$}}
\put(1370,613){\makebox(0,0)[r]{non-branching insertion sorting network \#mispredicts}}
\put(322,82){\makebox(0,0){$\circ$}}
\put(454,82){\makebox(0,0){$\circ$}}
\put(586,82){\makebox(0,0){$\circ$}}
\put(718,88){\makebox(0,0){$\circ$}}
\put(850,88){\makebox(0,0){$\circ$}}
\put(982,88){\makebox(0,0){$\circ$}}
\put(1114,88){\makebox(0,0){$\circ$}}
\put(1246,88){\makebox(0,0){$\circ$}}
\put(1378,88){\makebox(0,0){$\circ$}}
\put(1510,88){\makebox(0,0){$\circ$}}
\put(1642,88){\makebox(0,0){$\circ$}}
\put(1774,88){\makebox(0,0){$\circ$}}
\put(1906,88){\makebox(0,0){$\circ$}}
\put(1440,613){\makebox(0,0){$\circ$}}
\put(190.0,82.0){\rule[-0.200pt]{0.400pt}{187.179pt}}
\put(190.0,82.0){\rule[-0.200pt]{445.183pt}{0.400pt}}
\put(2038.0,82.0){\rule[-0.200pt]{0.400pt}{187.179pt}}
\put(190.0,859.0){\rule[-0.200pt]{445.183pt}{0.400pt}}
\end{picture}}
\caption{Comparing the number of branches, encountered and
  mispredicted, in optimized sorting algorithms for small numbers of
  inputs.}
\label{fig:branching}
\end{figure}

Figure~\ref{fig:branching} compares three sorting algorithms for small
numbers of inputs: (1) the unrolled insertion sort (also plotted in
Figure~\ref{fig:insertion}); (2) code derived from a standard
insertion sorting network (also plotted in
Figure~\ref{fig:insertion}); (3) the same insertion sorting network
but with a non-branching version of the \texttt{\color{magenta}COMP} macro. We
compare the number of branches encountered and mispredicted (averaged
over 100 million random executions).
From the figure it is clear that the number of branches encountered (and
mispredicted) is larger for both unrolled insertion sort and a naive
implementation of sorting networks. In contrast, the branching-free implementation
exhibits a nearly constant level of branches
encountered and mispredicted. These branches actually originate from
the surrounding test code (filling an array with random numbers,
computing random numbers, and checking that the result is actually
sorted).

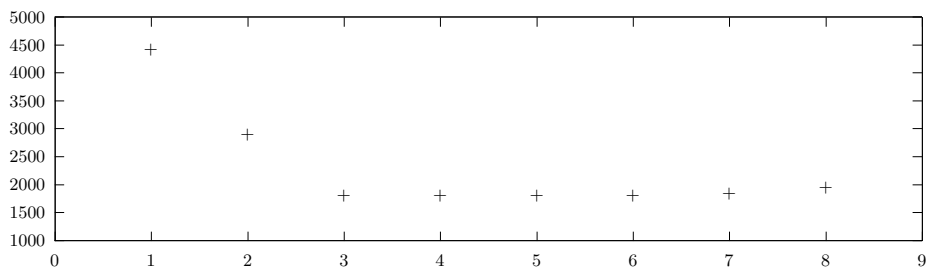
\begin{figure}[b]
\centering
\resizebox{\textwidth}{!}{
\setlength{\unitlength}{0.240900pt}
\ifx\plotpoint\undefined\newsavebox{\plotpoint}\fi
\sbox{\plotpoint}{\rule[-0.200pt]{0.400pt}{0.400pt}}%
\begin{picture}(1930,550)(130,50)
\sbox{\plotpoint}{\rule[-0.200pt]{0.400pt}{0.400pt}}%
\put(190.0,82.0){\rule[-0.200pt]{4.818pt}{0.400pt}}
\put(170,82){\makebox(0,0)[r]{$1000$}}
\put(2018.0,82.0){\rule[-0.200pt]{4.818pt}{0.400pt}}
\put(190.0,142.0){\rule[-0.200pt]{4.818pt}{0.400pt}}
\put(170,142){\makebox(0,0)[r]{$1500$}}
\put(2018.0,142.0){\rule[-0.200pt]{4.818pt}{0.400pt}}
\put(190.0,201.0){\rule[-0.200pt]{4.818pt}{0.400pt}}
\put(170,201){\makebox(0,0)[r]{$2000$}}
\put(2018.0,201.0){\rule[-0.200pt]{4.818pt}{0.400pt}}
\put(190.0,261.0){\rule[-0.200pt]{4.818pt}{0.400pt}}
\put(170,261){\makebox(0,0)[r]{$2500$}}
\put(2018.0,261.0){\rule[-0.200pt]{4.818pt}{0.400pt}}
\put(190.0,321.0){\rule[-0.200pt]{4.818pt}{0.400pt}}
\put(170,321){\makebox(0,0)[r]{$3000$}}
\put(2018.0,321.0){\rule[-0.200pt]{4.818pt}{0.400pt}}
\put(190.0,380.0){\rule[-0.200pt]{4.818pt}{0.400pt}}
\put(170,380){\makebox(0,0)[r]{$3500$}}
\put(2018.0,380.0){\rule[-0.200pt]{4.818pt}{0.400pt}}
\put(190.0,440.0){\rule[-0.200pt]{4.818pt}{0.400pt}}
\put(170,440){\makebox(0,0)[r]{$4000$}}
\put(2018.0,440.0){\rule[-0.200pt]{4.818pt}{0.400pt}}
\put(190.0,499.0){\rule[-0.200pt]{4.818pt}{0.400pt}}
\put(170,499){\makebox(0,0)[r]{$4500$}}
\put(2018.0,499.0){\rule[-0.200pt]{4.818pt}{0.400pt}}
\put(190.0,559.0){\rule[-0.200pt]{4.818pt}{0.400pt}}
\put(170,559){\makebox(0,0)[r]{$5000$}}
\put(2018.0,559.0){\rule[-0.200pt]{4.818pt}{0.400pt}}
\put(190.0,82.0){\rule[-0.200pt]{0.400pt}{4.818pt}}
\put(190,41){\makebox(0,0){$0$}}
\put(190.0,539.0){\rule[-0.200pt]{0.400pt}{4.818pt}}
\put(395.0,82.0){\rule[-0.200pt]{0.400pt}{4.818pt}}
\put(395,41){\makebox(0,0){$1$}}
\put(395.0,539.0){\rule[-0.200pt]{0.400pt}{4.818pt}}
\put(601.0,82.0){\rule[-0.200pt]{0.400pt}{4.818pt}}
\put(601,41){\makebox(0,0){$2$}}
\put(601.0,539.0){\rule[-0.200pt]{0.400pt}{4.818pt}}
\put(806.0,82.0){\rule[-0.200pt]{0.400pt}{4.818pt}}
\put(806,41){\makebox(0,0){$3$}}
\put(806.0,539.0){\rule[-0.200pt]{0.400pt}{4.818pt}}
\put(1011.0,82.0){\rule[-0.200pt]{0.400pt}{4.818pt}}
\put(1011,41){\makebox(0,0){$4$}}
\put(1011.0,539.0){\rule[-0.200pt]{0.400pt}{4.818pt}}
\put(1217.0,82.0){\rule[-0.200pt]{0.400pt}{4.818pt}}
\put(1217,41){\makebox(0,0){$5$}}
\put(1217.0,539.0){\rule[-0.200pt]{0.400pt}{4.818pt}}
\put(1422.0,82.0){\rule[-0.200pt]{0.400pt}{4.818pt}}
\put(1422,41){\makebox(0,0){$6$}}
\put(1422.0,539.0){\rule[-0.200pt]{0.400pt}{4.818pt}}
\put(1627.0,82.0){\rule[-0.200pt]{0.400pt}{4.818pt}}
\put(1627,41){\makebox(0,0){$7$}}
\put(1627.0,539.0){\rule[-0.200pt]{0.400pt}{4.818pt}}
\put(1833.0,82.0){\rule[-0.200pt]{0.400pt}{4.818pt}}
\put(1833,41){\makebox(0,0){$8$}}
\put(1833.0,539.0){\rule[-0.200pt]{0.400pt}{4.818pt}}
\put(2038.0,82.0){\rule[-0.200pt]{0.400pt}{4.818pt}}
\put(2038,41){\makebox(0,0){$9$}}
\put(2038.0,539.0){\rule[-0.200pt]{0.400pt}{4.818pt}}
\put(190.0,82.0){\rule[-0.200pt]{0.400pt}{114.909pt}}
\put(190.0,82.0){\rule[-0.200pt]{445.183pt}{0.400pt}}
\put(2038.0,82.0){\rule[-0.200pt]{0.400pt}{114.909pt}}
\put(190.0,559.0){\rule[-0.200pt]{445.183pt}{0.400pt}}
\put(395,490){\makebox(0,0){$+$}}
\put(601,309){\makebox(0,0){$+$}}
\put(806,178){\makebox(0,0){$+$}}
\put(1011,179){\makebox(0,0){$+$}}
\put(1217,178){\makebox(0,0){$+$}}
\put(1422,179){\makebox(0,0){$+$}}
\put(1627,182){\makebox(0,0){$+$}}
\put(1833,196){\makebox(0,0){$+$}}
\put(190.0,82.0){\rule[-0.200pt]{0.400pt}{114.909pt}}
\put(190.0,82.0){\rule[-0.200pt]{445.183pt}{0.400pt}}
\put(2038.0,82.0){\rule[-0.200pt]{0.400pt}{114.909pt}}
\put(190.0,559.0){\rule[-0.200pt]{445.183pt}{0.400pt}}
\end{picture}}
\caption{ILP on comparator networks of length $1000$ with differing levels of parallelism.}
\label{fig:ilp}
\end{figure}

Our second observation is that sorting networks are inherently
parallel, i.e., comparators at the same level can be performed simultaneously.
This
parallelism can be mapped directly to instruction level parallelism
(ILP).
The ability to
make use of ILP
has further performance potential.
In order to demonstrate this potential, we constructed artificial test
cases with varying levels of data dependency.
Given a natural number
$m$, we construct a comparator network of size $1000$ consisting of subsequences of $m$ parallel comparators. We would expect that, as $m$ grows,
we would see more use of ILP. 

In Figure~\ref{fig:ilp}, the values for $m$ are represented on the $x$-axis, while the $y$-axis (as usual) indicates the averaged number of CPU cycles. Indeed, we see significant performance
gains when going from $m = 1$ to $m = 2$ and $m = 3$. From this value onwards,
performance stays unchanged.
This is the result
of each comparator being compiled to $5$ assembler
instructions when using optimization level \texttt{-O3}. Then we
obtain slightly under $2$ CPU cycles per comparator.

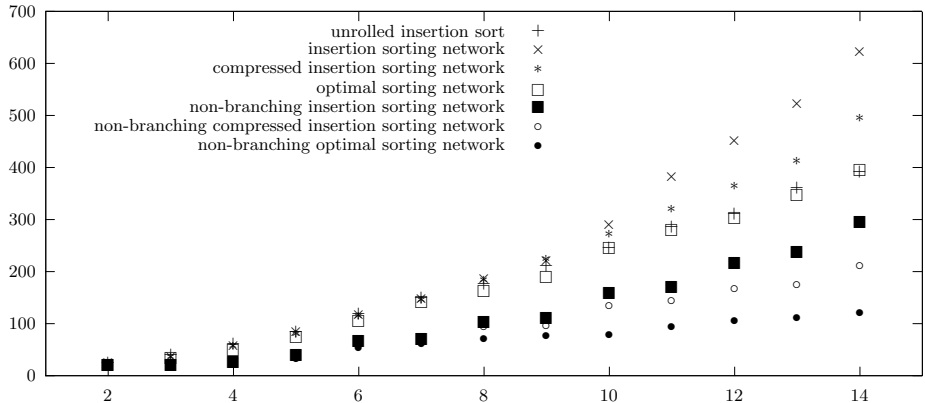
\begin{figure}[t]
\centering
\resizebox{\textwidth}{!}{
\setlength{\unitlength}{0.240900pt}
\ifx\plotpoint\undefined\newsavebox{\plotpoint}\fi
\sbox{\plotpoint}{\rule[-0.200pt]{0.400pt}{0.400pt}}%
\begin{picture}(1930,850)(130,50)
\sbox{\plotpoint}{\rule[-0.200pt]{0.400pt}{0.400pt}}%
\put(170.0,82.0){\rule[-0.200pt]{4.818pt}{0.400pt}}
\put(150,82){\makebox(0,0)[r]{$0$}}
\put(2018.0,82.0){\rule[-0.200pt]{4.818pt}{0.400pt}}
\put(170.0,193.0){\rule[-0.200pt]{4.818pt}{0.400pt}}
\put(150,193){\makebox(0,0)[r]{$100$}}
\put(2018.0,193.0){\rule[-0.200pt]{4.818pt}{0.400pt}}
\put(170.0,304.0){\rule[-0.200pt]{4.818pt}{0.400pt}}
\put(150,304){\makebox(0,0)[r]{$200$}}
\put(2018.0,304.0){\rule[-0.200pt]{4.818pt}{0.400pt}}
\put(170.0,415.0){\rule[-0.200pt]{4.818pt}{0.400pt}}
\put(150,415){\makebox(0,0)[r]{$300$}}
\put(2018.0,415.0){\rule[-0.200pt]{4.818pt}{0.400pt}}
\put(170.0,526.0){\rule[-0.200pt]{4.818pt}{0.400pt}}
\put(150,526){\makebox(0,0)[r]{$400$}}
\put(2018.0,526.0){\rule[-0.200pt]{4.818pt}{0.400pt}}
\put(170.0,637.0){\rule[-0.200pt]{4.818pt}{0.400pt}}
\put(150,637){\makebox(0,0)[r]{$500$}}
\put(2018.0,637.0){\rule[-0.200pt]{4.818pt}{0.400pt}}
\put(170.0,748.0){\rule[-0.200pt]{4.818pt}{0.400pt}}
\put(150,748){\makebox(0,0)[r]{$600$}}
\put(2018.0,748.0){\rule[-0.200pt]{4.818pt}{0.400pt}}
\put(170.0,859.0){\rule[-0.200pt]{4.818pt}{0.400pt}}
\put(150,859){\makebox(0,0)[r]{$700$}}
\put(2018.0,859.0){\rule[-0.200pt]{4.818pt}{0.400pt}}
\put(303.0,82.0){\rule[-0.200pt]{0.400pt}{4.818pt}}
\put(303,41){\makebox(0,0){$2$}}
\put(303.0,839.0){\rule[-0.200pt]{0.400pt}{4.818pt}}
\put(570.0,82.0){\rule[-0.200pt]{0.400pt}{4.818pt}}
\put(570,41){\makebox(0,0){$4$}}
\put(570.0,839.0){\rule[-0.200pt]{0.400pt}{4.818pt}}
\put(837.0,82.0){\rule[-0.200pt]{0.400pt}{4.818pt}}
\put(837,41){\makebox(0,0){$6$}}
\put(837.0,839.0){\rule[-0.200pt]{0.400pt}{4.818pt}}
\put(1104.0,82.0){\rule[-0.200pt]{0.400pt}{4.818pt}}
\put(1104,41){\makebox(0,0){$8$}}
\put(1104.0,839.0){\rule[-0.200pt]{0.400pt}{4.818pt}}
\put(1371.0,82.0){\rule[-0.200pt]{0.400pt}{4.818pt}}
\put(1371,41){\makebox(0,0){$10$}}
\put(1371.0,839.0){\rule[-0.200pt]{0.400pt}{4.818pt}}
\put(1638.0,82.0){\rule[-0.200pt]{0.400pt}{4.818pt}}
\put(1638,41){\makebox(0,0){$12$}}
\put(1638.0,839.0){\rule[-0.200pt]{0.400pt}{4.818pt}}
\put(1905.0,82.0){\rule[-0.200pt]{0.400pt}{4.818pt}}
\put(1905,41){\makebox(0,0){$14$}}
\put(1905.0,839.0){\rule[-0.200pt]{0.400pt}{4.818pt}}
\put(170.0,82.0){\rule[-0.200pt]{0.400pt}{187.179pt}}
\put(170.0,82.0){\rule[-0.200pt]{450.001pt}{0.400pt}}
\put(2038.0,82.0){\rule[-0.200pt]{0.400pt}{187.179pt}}
\put(170.0,859.0){\rule[-0.200pt]{450.001pt}{0.400pt}}
\put(1150,818){\makebox(0,0)[r]{unrolled insertion sort}}
\put(303,109){\makebox(0,0){$+$}}
\put(437,127){\makebox(0,0){$+$}}
\put(570,150){\makebox(0,0){$+$}}
\put(704,177){\makebox(0,0){$+$}}
\put(837,215){\makebox(0,0){$+$}}
\put(971,248){\makebox(0,0){$+$}}
\put(1104,278){\makebox(0,0){$+$}}
\put(1237,316){\makebox(0,0){$+$}}
\put(1371,355){\makebox(0,0){$+$}}
\put(1504,401){\makebox(0,0){$+$}}
\put(1638,427){\makebox(0,0){$+$}}
\put(1771,483){\makebox(0,0){$+$}}
\put(1905,517){\makebox(0,0){$+$}}
\put(1220,818){\makebox(0,0){$+$}}
\put(1150,777){\makebox(0,0)[r]{insertion sorting network}}
\put(303,109){\makebox(0,0){$\times$}}
\put(437,123){\makebox(0,0){$\times$}}
\put(570,147){\makebox(0,0){$\times$}}
\put(704,175){\makebox(0,0){$\times$}}
\put(837,212){\makebox(0,0){$\times$}}
\put(971,247){\makebox(0,0){$\times$}}
\put(1104,289){\makebox(0,0){$\times$}}
\put(1237,328){\makebox(0,0){$\times$}}
\put(1371,405){\makebox(0,0){$\times$}}
\put(1504,507){\makebox(0,0){$\times$}}
\put(1638,584){\makebox(0,0){$\times$}}
\put(1771,663){\makebox(0,0){$\times$}}
\put(1905,773){\makebox(0,0){$\times$}}
\put(1220,777){\makebox(0,0){$\times$}}
\put(1150,736){\makebox(0,0)[r]{compressed insertion sorting network}}
\put(303,108){\makebox(0,0){$\ast$}}
\put(437,122){\makebox(0,0){$\ast$}}
\put(570,145){\makebox(0,0){$\ast$}}
\put(704,172){\makebox(0,0){$\ast$}}
\put(837,209){\makebox(0,0){$\ast$}}
\put(971,244){\makebox(0,0){$\ast$}}
\put(1104,286){\makebox(0,0){$\ast$}}
\put(1237,331){\makebox(0,0){$\ast$}}
\put(1371,384){\makebox(0,0){$\ast$}}
\put(1504,438){\makebox(0,0){$\ast$}}
\put(1638,487){\makebox(0,0){$\ast$}}
\put(1771,540){\makebox(0,0){$\ast$}}
\put(1905,632){\makebox(0,0){$\ast$}}
\put(1220,736){\makebox(0,0){$\ast$}}
\put(1150,695){\makebox(0,0)[r]{optimal sorting network}}
\put(303,109){\raisebox{-.8pt}{\makebox(0,0){$\Box$}}}
\put(437,122){\raisebox{-.8pt}{\makebox(0,0){$\Box$}}}
\put(570,141){\raisebox{-.8pt}{\makebox(0,0){$\Box$}}}
\put(704,168){\raisebox{-.8pt}{\makebox(0,0){$\Box$}}}
\put(837,201){\raisebox{-.8pt}{\makebox(0,0){$\Box$}}}
\put(971,242){\raisebox{-.8pt}{\makebox(0,0){$\Box$}}}
\put(1104,265){\raisebox{-.8pt}{\makebox(0,0){$\Box$}}}
\put(1237,296){\raisebox{-.8pt}{\makebox(0,0){$\Box$}}}
\put(1371,357){\raisebox{-.8pt}{\makebox(0,0){$\Box$}}}
\put(1504,395){\raisebox{-.8pt}{\makebox(0,0){$\Box$}}}
\put(1638,421){\raisebox{-.8pt}{\makebox(0,0){$\Box$}}}
\put(1771,471){\raisebox{-.8pt}{\makebox(0,0){$\Box$}}}
\put(1905,523){\raisebox{-.8pt}{\makebox(0,0){$\Box$}}}
\put(1220,695){\raisebox{-.8pt}{\makebox(0,0){$\Box$}}}
\put(1150,654){\makebox(0,0)[r]{non-branching insertion sorting network}}
\put(303,105){\makebox(0,0){$\blacksquare$}}
\put(437,105){\makebox(0,0){$\blacksquare$}}
\put(570,111){\makebox(0,0){$\blacksquare$}}
\put(704,125){\makebox(0,0){$\blacksquare$}}
\put(837,156){\makebox(0,0){$\blacksquare$}}
\put(971,160){\makebox(0,0){$\blacksquare$}}
\put(1104,197){\makebox(0,0){$\blacksquare$}}
\put(1237,204){\makebox(0,0){$\blacksquare$}}
\put(1371,259){\makebox(0,0){$\blacksquare$}}
\put(1504,272){\makebox(0,0){$\blacksquare$}}
\put(1638,323){\makebox(0,0){$\blacksquare$}}
\put(1771,345){\makebox(0,0){$\blacksquare$}}
\put(1905,410){\makebox(0,0){$\blacksquare$}}
\put(1220,654){\makebox(0,0){$\blacksquare$}}
\put(1150,613){\makebox(0,0)[r]{non-branching compressed insertion sorting network}}
\put(303,105){\makebox(0,0){$\circ$}}
\put(437,105){\makebox(0,0){$\circ$}}
\put(570,112){\makebox(0,0){$\circ$}}
\put(704,124){\makebox(0,0){$\circ$}}
\put(837,155){\makebox(0,0){$\circ$}}
\put(971,153){\makebox(0,0){$\circ$}}
\put(1104,185){\makebox(0,0){$\circ$}}
\put(1237,187){\makebox(0,0){$\circ$}}
\put(1371,230){\makebox(0,0){$\circ$}}
\put(1504,241){\makebox(0,0){$\circ$}}
\put(1638,266){\makebox(0,0){$\circ$}}
\put(1771,275){\makebox(0,0){$\circ$}}
\put(1905,316){\makebox(0,0){$\circ$}}
\put(1220,613){\makebox(0,0){$\circ$}}
\put(1150,572){\makebox(0,0)[r]{non-branching optimal sorting network}}
\put(303,105){\makebox(0,0){$\bullet$}}
\put(437,106){\makebox(0,0){$\bullet$}}
\put(570,108){\makebox(0,0){$\bullet$}}
\put(704,118){\makebox(0,0){$\bullet$}}
\put(837,142){\makebox(0,0){$\bullet$}}
\put(971,149){\makebox(0,0){$\bullet$}}
\put(1104,161){\makebox(0,0){$\bullet$}}
\put(1237,166){\makebox(0,0){$\bullet$}}
\put(1371,169){\makebox(0,0){$\bullet$}}
\put(1504,186){\makebox(0,0){$\bullet$}}
\put(1638,198){\makebox(0,0){$\bullet$}}
\put(1771,204){\makebox(0,0){$\bullet$}}
\put(1905,215){\makebox(0,0){$\bullet$}}
\put(1220,572){\makebox(0,0){$\bullet$}}
\put(170.0,82.0){\rule[-0.200pt]{0.400pt}{187.179pt}}
\put(170.0,82.0){\rule[-0.200pt]{450.001pt}{0.400pt}}
\put(2038.0,82.0){\rule[-0.200pt]{0.400pt}{187.179pt}}
\put(170.0,859.0){\rule[-0.200pt]{450.001pt}{0.400pt}}
\end{picture}}
\caption{Comparison of sorting networks for small numbers of inputs:
  non-branching sorting networks are fastest.}
\label{fig:implementation}
\end{figure}

Combining the gains from ILP with the absence of branching, we obtain
large speed-ups for small inputs when comparing to both insertion sort
and naive implementations of sorting networks. In
Figure~\ref{fig:implementation}, we show the magnitude of the
improvements obtained. Once again we plot the number of inputs
on the $x$-axis against the number of cycles required to sort then
on the $y$-axis, averaged over $100$ million random executions. We
consider the unrolled insertion sort, the three sorting networks from
Figure~\ref{fig:insertion} (insertion sorting network, compressed
insertion sorting network, and optimal sorting network), and 
these same three sorting networks using non-branching
comparators (non-branching insertion sorting network, non-branching compressed insertion
sorting network, and non-branching optimal sorting network).
The figure shows that using the best known (optimal) sorting
networks in their non-branching forms results in a speed-up by a factor of more than $3$.

\section{Quicksort with Sorting Network Base Case}
\label{sec:quicksort}

We now demonstrate that optimizing the code in the base case of a
Quicksort algorithm translates to real-world savings when applying the
sorting function. To this end, we use as base cases (1) the
(empirically) best variant of insertion sort unrolled by applying
program transformations to the algorithm from~\cite{SedgewickBook}, and (2) the
fastest non-branching code derived from optimal (size) sorting
networks.

\begin{figure}[t]
\centering
\resizebox{\textwidth}{!}{
\setlength{\unitlength}{0.240900pt}
\ifx\plotpoint\undefined\newsavebox{\plotpoint}\fi
\sbox{\plotpoint}{\rule[-0.200pt]{0.400pt}{0.400pt}}%
\begin{picture}(1930,850)(130,50)
\sbox{\plotpoint}{\rule[-0.200pt]{0.400pt}{0.400pt}}%
\put(210.0,82.0){\rule[-0.200pt]{4.818pt}{0.400pt}}
\put(190,82){\makebox(0,0)[r]{$110{,}000$}}
\put(2018.0,82.0){\rule[-0.200pt]{4.818pt}{0.400pt}}
\put(210.0,212.0){\rule[-0.200pt]{4.818pt}{0.400pt}}
\put(190,212){\makebox(0,0)[r]{$120{,}000$}}
\put(2018.0,212.0){\rule[-0.200pt]{4.818pt}{0.400pt}}
\put(210.0,341.0){\rule[-0.200pt]{4.818pt}{0.400pt}}
\put(190,341){\makebox(0,0)[r]{$130{,}000$}}
\put(2018.0,341.0){\rule[-0.200pt]{4.818pt}{0.400pt}}
\put(210.0,471.0){\rule[-0.200pt]{4.818pt}{0.400pt}}
\put(190,471){\makebox(0,0)[r]{$140{,}000$}}
\put(2018.0,471.0){\rule[-0.200pt]{4.818pt}{0.400pt}}
\put(210.0,600.0){\rule[-0.200pt]{4.818pt}{0.400pt}}
\put(190,600){\makebox(0,0)[r]{$150{,}000$}}
\put(2018.0,600.0){\rule[-0.200pt]{4.818pt}{0.400pt}}
\put(210.0,730.0){\rule[-0.200pt]{4.818pt}{0.400pt}}
\put(190,730){\makebox(0,0)[r]{$160{,}000$}}
\put(2018.0,730.0){\rule[-0.200pt]{4.818pt}{0.400pt}}
\put(210.0,859.0){\rule[-0.200pt]{4.818pt}{0.400pt}}
\put(190,859){\makebox(0,0)[r]{$170{,}000$}}
\put(2018.0,859.0){\rule[-0.200pt]{4.818pt}{0.400pt}}
\put(341.0,82.0){\rule[-0.200pt]{0.400pt}{4.818pt}}
\put(341,41){\makebox(0,0){$2$}}
\put(341.0,839.0){\rule[-0.200pt]{0.400pt}{4.818pt}}
\put(602.0,82.0){\rule[-0.200pt]{0.400pt}{4.818pt}}
\put(602,41){\makebox(0,0){$4$}}
\put(602.0,839.0){\rule[-0.200pt]{0.400pt}{4.818pt}}
\put(863.0,82.0){\rule[-0.200pt]{0.400pt}{4.818pt}}
\put(863,41){\makebox(0,0){$6$}}
\put(863.0,839.0){\rule[-0.200pt]{0.400pt}{4.818pt}}
\put(1124.0,82.0){\rule[-0.200pt]{0.400pt}{4.818pt}}
\put(1124,41){\makebox(0,0){$8$}}
\put(1124.0,839.0){\rule[-0.200pt]{0.400pt}{4.818pt}}
\put(1385.0,82.0){\rule[-0.200pt]{0.400pt}{4.818pt}}
\put(1385,41){\makebox(0,0){$10$}}
\put(1385.0,839.0){\rule[-0.200pt]{0.400pt}{4.818pt}}
\put(1646.0,82.0){\rule[-0.200pt]{0.400pt}{4.818pt}}
\put(1646,41){\makebox(0,0){$12$}}
\put(1646.0,839.0){\rule[-0.200pt]{0.400pt}{4.818pt}}
\put(1907.0,82.0){\rule[-0.200pt]{0.400pt}{4.818pt}}
\put(1907,41){\makebox(0,0){$14$}}
\put(1907.0,839.0){\rule[-0.200pt]{0.400pt}{4.818pt}}
\put(210.0,82.0){\rule[-0.200pt]{0.400pt}{187.179pt}}
\put(210.0,82.0){\rule[-0.200pt]{440.365pt}{0.400pt}}
\put(2038.0,82.0){\rule[-0.200pt]{0.400pt}{187.179pt}}
\put(210.0,859.0){\rule[-0.200pt]{440.365pt}{0.400pt}}
\put(1110,818){\makebox(0,0)[r]{unrolled insertion sort}}
\put(341,628){\makebox(0,0){$+$}}
\put(471,597){\makebox(0,0){$+$}}
\put(602,572){\makebox(0,0){$+$}}
\put(732,539){\makebox(0,0){$+$}}
\put(863,507){\makebox(0,0){$+$}}
\put(993,477){\makebox(0,0){$+$}}
\put(1124,470){\makebox(0,0){$+$}}
\put(1255,445){\makebox(0,0){$+$}}
\put(1385,427){\makebox(0,0){$+$}}
\put(1516,422){\makebox(0,0){$+$}}
\put(1646,399){\makebox(0,0){$+$}}
\put(1777,394){\makebox(0,0){$+$}}
\put(1907,397){\makebox(0,0){$+$}}
\put(1180,818){\makebox(0,0){$+$}}
\put(1110,777){\makebox(0,0)[r]{non-branching optimal sorting network}}
\put(341,593){\makebox(0,0){$\times$}}
\put(471,526){\makebox(0,0){$\times$}}
\put(602,486){\makebox(0,0){$\times$}}
\put(732,425){\makebox(0,0){$\times$}}
\put(863,389){\makebox(0,0){$\times$}}
\put(993,361){\makebox(0,0){$\times$}}
\put(1124,320){\makebox(0,0){$\times$}}
\put(1255,278){\makebox(0,0){$\times$}}
\put(1385,244){\makebox(0,0){$\times$}}
\put(1516,232){\makebox(0,0){$\times$}}
\put(1646,212){\makebox(0,0){$\times$}}
\put(1777,187){\makebox(0,0){$\times$}}
\put(1907,178){\makebox(0,0){$\times$}}
\put(1180,777){\makebox(0,0){$\times$}}
\put(210.0,82.0){\rule[-0.200pt]{0.400pt}{187.179pt}}
\put(210.0,82.0){\rule[-0.200pt]{440.365pt}{0.400pt}}
\put(2038.0,82.0){\rule[-0.200pt]{0.400pt}{187.179pt}}
\put(210.0,859.0){\rule[-0.200pt]{440.365pt}{0.400pt}}
\end{picture}}
\caption{Quicksort: comparing insertion sort at the base case with
  non-branching optimal sorting networks at the base case. Plotting
  base case size ($x$ axis) and number of cycles (averaged over one 
  million random runs).}
\label{fig:qsort}
\end{figure}
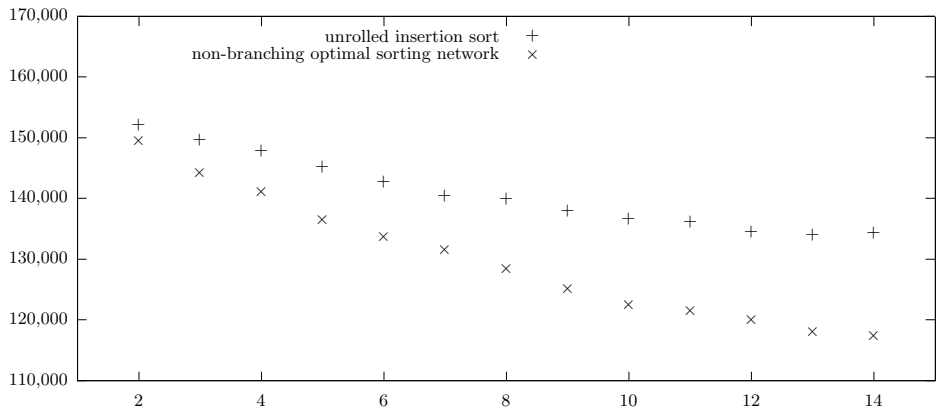

In Figure~\ref{fig:qsort} we depict the results of sorting
lists of $10{,}000$ elements.
The $y$-axis measures the number of
cycles (averaged over one million random runs), and the $x$-axis specifies the
limit at which Quicksort reverts to a base case. For example, the
value $8$ indicates that the algorithm uses a base case whenever it is
required to sort a sequence of length \emph{at most} $8$ elements. The value
$2$ corresponds to the case where the base case has no impact.
To quantify the impact of the choice of base case, we compare to the
case for value $2$ (on the $x$-axis).  For insertion sort we see a $2$--$12\%$
reduction in runtime depending on
the limit, and for non-branching sorting networks we
achieve instead  $7$--$23\%$ reduction in runtime.

\section{Conclusion}
\label{sec:conclusion}
In this paper, we showed, both theoretically and empirically, that using
code derived naively from sorting networks is not advantageous
to sort small numbers of inputs, compared to the use of standard
data-dependent sorting algorithms like insertion
sort. Furthermore, we showed that program transformations are of only
limited utility for improving insertion sort on small numbers of
inputs.

By contrast, we showed how to synthesize simple yet efficient implementations of sorting
networks, and gave insight into the microarchitectural features that
enable this implementation. We demonstrated that we do obtain significant speed-ups
compared to naive implementations such as \cite{Lopez2014}. A further empirical comparison
between our implementation and the one described in \cite{Furtak2007}
(not detailed in this paper) shows
similar performance and scaling behavior.
However, our approach allows the exploitation of instruction-level parallelism
without the need for a complex instruction set-specific algorithm, as required by \cite{Furtak2007}.
We also provided further evidence that efficient sorting networks are useful as
a base case in divide-and-conquer sorting algorithms such as, e.g., Quicksort.

Our results also show that using different
sorting networks has measurable impact on the efficiency of the
synthesized C code. While previous research on finding optimal
sorting networks has focused on optimal depth or optimal size, in the future we plan
to identify criteria that will lead to optimal performance in this
context. What are the parameters that determine real-world
efficiency of the synthesized code, and how can we find
sorting networks that optimize these parameters?
We also plan to explore other target architectures, such as GPUs,
and to benchmark our approach as base case for other sorting algorithms, such as merge sort.

\bibliographystyle{plain}
\bibliography{paper}

\end{document}